\documentclass[reprint,amsmath,amssymb,aps,superscriptaddress,pra]{revtex4-1}

\usepackage[utf8]{inputenc}
\usepackage[english]{babel}
\usepackage[english]{babelbib}
\usepackage{amsmath,amssymb}
\usepackage{braket}
\usepackage{bm}
\usepackage{graphicx}
\graphicspath{{figuras/}}
\usepackage{soul}
\usepackage{float}
\usepackage{verbatim}
\usepackage[normalem]{ulem}
\usepackage{xcolor}
\usepackage[dvipsnames]{xcolor}
\usepackage[pdftex,colorlinks=true,urlcolor= blue,linkcolor=Blue,citecolor=BrickRed]{hyperref}
\usepackage{cases} 
\usepackage{media9}
\usepackage{animate}
\usepackage{epstopdf}
\AppendGraphicsExtensions{.gif}
\usepackage{braket}
\usepackage[normalem]{ulem}
\usepackage{soul}

\def\PT{\mathcal{PT}}
\def\mcpt{\mathcal{PT}}

\def\HC{H^{\dagger}}

\newcommand{\steH}{\widetilde{\varphi}}
\newcommand{\steHC}{\overline{\psi}}

\newcommand{\steini}{\varphi}
\newcommand{\stang}{\phi}

\newcommand{\ketH}[1]{\ket{\steH_{#1}}}
\newcommand{\ketHC}[1]{\ket{\steHC_{#1}}}

\newcommand{\braHC}[1]{\bra{\steHC_{#1}}}

\newcommand{\braketbar}[2]{\langle{#1}|{#2}\rangle}
\newcommand{\re}{\mathrm{e}}
\def\th{\theta}

\def\al{\alpha}
\def\be{\beta}
\def\sg{\sigma}

\def\uni{{\bf i}}

\def\beqn{\begin{eqnarray}}
\def\eeqn{\end{eqnarray}}
\def\nonn{\nonumber\\}
\def\wid{\widetilde}
\def\ovl{\overline}
\def\nonn{\nonumber \\}
\def\la{\langle}
\def\ra{\rangle}

\newcommand{\CC}{\mathbb{C}}

\newcommand{\MR}[1]{{ #1}}

\begin{document}


\title{Uncertainty inequalities in a non-Hermitian scenario }


\newcommand{\AffilA}{Instituto de F\'isica La Plata (IFLP), CONICET--UNLP \ and \ Facultad de Ciencias Exactas, Universidad Nacional de La Plata, 1900 La Plata, Argentina}   
\newcommand{\AffilB}{Instituto Argentino de Matemática (IAM), CONICET \ and \ CMaLP,  Facultad de Ciencias Exactas, Universidad Nacional de La Plata, 1900 La Plata, Argentina}

\author{Yanet {Alvarez}}
\email{yalvarez@fisica.unlp.edu.ar}
\affiliation{\AffilA}
\author{Mariela Portesi}
\email{portesi@fisica.unlp.edu.ar}
\affiliation{\AffilA}
\author{Romina Ramírez}
\email{romina@mate.unlp.edu.ar}
\affiliation{\AffilB}
%
\author{Marta Reboiro}%
\email{reboiro@fisica.unlp.edu.ar}
\affiliation{\AffilA}

\date{\today}

\begin{abstract}
We investigate uncertainty relations for quantum observables evolving under non-Hermitian Hamiltonians, with particular emphasis on the role of metric operators. By constructing appropriate metrics in each dynamical regime, namely the unbroken-symmetry phase, the spontaneously broken-symmetry phase, and at exceptional points, we provide a consistent definition of expectation values, variances, and time evolution within a Krein-space framework. Within this approach, we derive a generalized Heisenberg–Robertson uncertainty inequality  
which is valid across all spectral regimes. As an application, we analyze a spin 
model with parity-time reversal symmetry
and show that, while the uncertainty measure exhibits oscillatory behavior in the unbroken phase, it evolves towards a minimum-uncertainty steady state in the spontaneously broken-symmetry phase and at exceptional points. We further compare our metric-based description with a Lindblad master-equation approach and show their agreement in the steady state. Our results highlight the necessity of incorporating appropriate metric structures to extract physically meaningful predictions from non-Hermitian quantum dynamics.
\end{abstract}             
          
\maketitle

\section{Introduction}
\label{intro}

Non-Hermitian Hamiltonians have become a widely used framework for describing effective quantum dynamics in open, driven, and dissipative systems \cite{Ben98,BenderBrodyJones2002,el2018non,ashida2020non,bendereview}. In particular, Hamiltonians possessing parity–time reversal ($\PT$) symmetry provide experimentally relevant models exhibiting real spectra, dynamical phase transitions, and exceptional points \cite{pastawski,epjd2020,swanson,berg,xue1}. Despite their success in capturing observable phenomena, non-Hermitian formulations raise fundamental questions concerning the definition of physical observables, expectation values, and quantum uncertainties, especially when the standard inner product of quantum mechanics is no longer adequate \cite{Scholtz1992,bio,BenderBrodyJones2002,brody2013biorthogonal,Mostafazadeh2006Cubic,Mostafazadeh2007TimeDep,Znojil2007,znojil2024non,Bagarello2015Book,pseudo1,fringtmetric,Ram19,mandel,Bag23,znojil2025}.  
  
Recent developments in quantum theory highlight the growing relevance of uncertainty relations beyond the standard Hermitian framework. In particular, uncertainty relations for non-Hermitian models have been used to construct generalized bounds in quantum metrology, such as modified Cramér–Rao inequalities \cite{cramerrao}, revealing new connections between measurement precision and non-Hermitian dynamics. These advances raise important questions regarding how non-Hermiticity affects foundational and operational aspects of conventional uncertainty relations. For instance, investigations of nonclassicality based on uncertainty constraints such as those explored in Ref.~\cite{spekkens} as well as recent developments in stronger formulations of uncertainty relations \cite{maccone}, may acquire new structure when generalized to non-Hermitian settings. The formalism developed in this work naturally lends itself to addressing such questions, suggesting potential extensions of established uncertainty-based frameworks and thereby broadening their applicability in contemporary quantum theory.

A consistent quantum-mechanical interpretation of non-Hermitian dynamics requires the introduction of a modified inner product defined through an appropriate metric operator \cite{bio,BenderBrodyJones2002,Ben12,brody2013biorthogonal,Mostafazadeh2006Cubic,Mostafazadeh2007TimeDep,Znojil2007,znojil2024non,Bagarello2015Book,
pseudo1,fringtmetric,Ram19,mandel,znojil2025}. For pseudo-Hermitian Hamiltonians \cite{pseudo1}, such a metric renders the Hamiltonian self-adjoint with respect to the modified inner product \cite{pseudo1,Mostafazadeh2008MetricOps,Mostafazadeh2006Cubic}. While this construction is well understood in the unbroken-symmetry regime, where the spectrum is entirely real, its extension beyond this phase remains subtle. In the spontaneously broken-symmetry regime, where pairs of complex-conjugate eigenvalues appear, and at exceptional points, where eigenvectors coalesce, the metric may become indefinite or singular, therefore a consistent definition of expectation values of observables 
is not straightforward 
\cite{Bagarello2015Book,Ram19,Bag23,mandel}.   
This has direct implications in the establishment of appropriate uncertainty relations (see, for instance, \cite{Hei27,Rob29,urgen1,urgen2,urgen3,ozawa}  and refs.\ therein).

   
In the present contribution,  
we address this problem by constructing metric operators adapted to each dynamical regime for  
a finite-dimensional pseudo-Hermitian Hamiltonian: the unbroken-symmetry phase, the {{spontaneously}} broken-symmetry phase, and the exceptional points \cite{Ram19}. Our approach  makes use of the formalism of Krein-spaces \cite{krein,krein1,albeverio2004,Bagarello2015Book}, which allow for a consistent treatment of indefinite metrics and provides a unified definition of expectation values, variances, and time evolution across all spectral regimes. Within this formulation, we derive a generalized Heisenberg-–Robertson uncertainty relation that remains valid irrespective of the spectral properties of the Hamiltonian.

As an explicit application, we analyze a $\PT$-symmetric spin system and compute the uncertainty relations associated with non-commuting spin observables. We show that, in the unbroken-symmetry phase, an oscillatory behavior in time appears, 
while in the {{spontaneously}} broken-symmetry phase and at exceptional points the system 
evolves towards a minimum-uncertainty steady state. 

Finally, we compare our metric-based description,{{ for the spontaneously broken symmetry phase}}, with a Lindblad master-equation approach for open quantum systems \cite{lindblad1976generators,nori,nathan2020universal}. We show that both frameworks lead to the same steady-state behavior, thereby establishing the physical consistency of our construction and clarifying the role of metric operators in extracting meaningful dynamical information from non-Hermitian quantum systems beyond the exact symmetry phase.
Moreover, we show that in the {{spontaneously}} broken symmetry phase the eigenfunctions of the studied Hamiltonian are not orthogonal with respect to the inner product we have introduced, which is in accordance with 
recent experimental results on dissipative systems \cite{joglekar2019}.

The article
is organized as follows.
In Section~\ref{formalismo}, we present the procedure to construct a suitable metric operator $S$ for each spectral regime: the symmetric phase in \ref{sub211}, the {{spontaneously}} broken-symmetry phase in~\ref{sub212}, and the exceptional points in~\ref{sub213}. 
The observables' mean values are given in \ref{sub22}.
In Subsection~\ref{time_evol}, we further construct the time evolution of the initial state together with the corresponding expectation values. Section~\ref{results} is devoted to compute the modified Heisenberg–-Robertson uncertainty relations for a pseudo-Hermitian $\mathcal{PT}$-symmetric Hamiltonian model, focusing on the observables $\sigma_x$ and $\sigma_y$. First, we verify this inequality for a general state. Finally, we extend the
analysis to the time evolution of a given initial state.
These calculations are carried out across all spectral regimes, including the exceptional points. 
We compare our results with those obtained using the Lindblad formalism.
Concluding remarks are drawn in Section~\ref{conclusions}.

\section{Formalism}
\label{formalismo}


As it is well known, a Hamiltonian $H$ is said to be \emph{pseudo-Hermitian}~\cite{pseudo1} in a Hilbert space $\mathcal{H}$ if there exists an invertible Hermitian operator $S$ such that 
\beqn
H^\dagger S = S H,
\label{pseudo}
\eeqn
in a dense domain $\mathcal{D}(H)$. 
As a consequence of the similarity relation between $H$ and $H^\dagger$, 
the spectrum of $H$ is either real or contains complex conjugate pairs of eigenvalues. 
This property gives rise to the so-called \emph{dynamical phase transition} \cite{pastawski} 
in the parameter space of the model. 
In the \emph{unbroken-symmetry phase}, the eigenvectors of $H$ respect the symmetry of the Hamiltonian, 
and the spectrum is entirely real. 
In the \emph{{{spontaneously}} broken-symmetry phase}, the spectrum contains complex conjugate pairs of eigenvalues, 
and the corresponding eigenvectors no longer preserve the symmetry of the Hamiltonian. 
The boundary between these two phases consists of \emph{exceptional points (EPs)}, 
where two or more eigenvalues coalesce and their associated eigenvectors become linearly dependent.

The analysis of non-Hermitian Hamiltonians endowed with modified inner products has a long history, including both the quasi-Hermitian and pseudo-Hermitian frameworks and earlier indefinite-metric approaches \cite{Bagarello2015Book}. In $\mathcal{PT}$-symmetric models, the familiar constructions based on a $\mathcal{C}$-type operator in the unbroken-symmetry regime can be viewed as particular realizations of this broader metric-operator program \cite{BenderBrodyJones2002,Ben12}. 
A review of pseudo-Hermitian quantum mechanics and its physical interpretation, including the construction of similarity relations between a Hamiltonian and its adjoint can be found in \cite{pseudo1}. Its main purpose is to explain under which conditions a non-Hermitian Hamiltonian can still define a consistent quantum theory, namely when a positive-definite metric operator can be introduced so that the observable operators become Hermitian with respect to the physical inner product. In this sense, in Ref.~\cite{pseudo1} PT-symmetric models are considered within the broader framework of pseudo-Hermitian and quasi-Hermitian quantum mechanics, and emphasizes that the metric operator, rather than PT symmetry alone, is the central object in the construction.
The scope of the article is mainly the quasi-Hermitian regime where a unitary formulation is possible. It also discusses how observables are defined in the new physical Hilbert space, and how the non-Hermitian description is related to an equivalent Hermitian one through similarity transformations; in addition, these ideas are illustrated with analytical models, and broader issues such as time-dependent Hamiltonians, path-integral methods, and applications are also commented.

In Refs.~\cite{fringtmetric,fringtm1,fring2023} and references therein, time-dependent metrics are constructed both for \MR{time-dependent and time-independent Hamiltonians}. The authors show that this metric is valid at the exceptional points, as well as in the broken and the unbroken \MR{regimes}.

\MR{It is also worth mentioning that in~\cite{Ram19}} we have extended the previous time-independent approaches to construct physical Hilbert spaces for non-\MR{Hermitian} general Hamiltonians. The formalism includes the case of pseudo-\MR{Hermitian} Hamiltonians, and it is valid beyond the quasi-\MR{Hermitian} regime and at  exceptional points. 

Recently, the authors of Ref.~\cite{mandel} have analyzed uncertainty relations of a pseudo-Hermitian Hamiltonian. They have shown that, although a Hamiltonian may be pseudo-Hermitian, the associated similarity transformation does not necessarily yield a consistent inner product, and thus cannot always be used to define expectation values of observables.

  
In order to develop a physically consistent formulation of quantum uncertainty relations for non-commuting observables in systems governed by non-Hermitian Hamiltonian dynamics, we 
make use here of 
the formalism introduced in Ref.~\cite{Ram19}. The essential elements of this formalism can be summarized as follows:
(i)~the construction of metric operators within the framework of Krein spaces in order to define a new inner product, and
(ii)~the transformation of operators associated with observables so that they become self-adjoint with respect to the inner product induced by the metric operator.

Let us briefly review the essentials of our proposal.

\subsection{Metric operators} 
\label{subs21}

Let $\ketH{\alpha}$ and $\ketHC{\beta}$ be eigenvectors of $H$ and $H^\dagger$, respectively. That is
\begin{eqnarray}
H\ketH{\alpha} & = & \wid E_\alpha\ketH{\alpha},\nonn 
H^\dagger\ketHC{\beta} & = &\ovl E_\beta \ketHC{\beta}.
\end{eqnarray}
Due to Eq.~\eqref{pseudo}, $S \ketH{\al}$ is an eigenvector of $H^\dag$ with eigenvalue $\wid E_\al$.

It is well established that the eigenvectors of $H$ and $H^\dagger$ form a complete biorthogonal set \cite{bio,pseudo1}

\beqn
\langle \ovl \psi_\be | \wid \varphi_\al \rangle =\langle S \wid \varphi_\be | \wid \varphi_\al \rangle=\langle  \wid \varphi_\be | \wid \varphi_\al \rangle_S= ~\delta_{\al,\be},~~~\ovl E_\beta= \wid E_\alpha^*.
\nonn
\eeqn

If $S$ is positive definite, a new Hilbert space endowed with the inner product 
$\langle \cdot | \cdot \rangle_S$ can be introduced. As that is not always the case, we proceed as follows, depending on the spectral region.

\subsubsection{Symmetry phase}
\label{sub211}

Following Ref.~\cite{Ram19}, we take as metric the operator $S$ given by
\beqn
S = \sum_\al \ketHC{\al} \braHC{\al}.
\label{spt}
\eeqn
Since
$S$ is Hermitian and positive definite, it can be written in the form 
$S=\Upsilon^\dagger \Upsilon$. 
We construct the
Hilbert space by introducing a new inner product induced by the metric operator $S$, \ 
$\langle . | . \rangle_S = \langle S . | . \rangle$.

It should be noticed that $S$ is a similarity operator, $H^\dag S = S H$. Moreover, as it is positive definite, $H$ and $H^\dag$ are  similar to a Hermitian Hamiltonian $h$:
\beqn
h= \Upsilon H \Upsilon^{-1}= \Upsilon^{\dag ~-1} H^\dag \Upsilon^\dag.
\eeqn

\subsubsection{{{Spontaneously}} broken symmetry phase}
\label{sub212}

In the {{spontaneously}} broken symmetry phase, one can 
take as a similarity operator between $H$ and $H^\dag$, the operator 
\beqn
    S = \sum_{\al \le \be}^{N} ~ \delta(\ovl{E}_\be-\wid{E}_\al^*) \left ( \zeta_\be \ketHC{\be} \braHC{\al}+ \zeta_\be^* \ketHC{\al} \braHC{\be} \right),
\label{snpt}
\eeqn
with $\zeta \in \CC$ and $Im(\zeta) \neq 0$ being $N$ the dimension of the Hilbert space.

So defined, it is easy to prove that $H^\dag S = S H$. However, $S$ in Eq.~\eqref{snpt} is in general non-positive definite. Thus, $S$ induces an indefinite
metric. To overcome this difficulty, we suggest as strategic to adopt the formalism of Krein spaces \cite{krein}. 

It is possible to construct a basis of the Hilbert space, $\cal{B}_S$, with the eigenvectors of $S$. This basis can be split into two sets by separating the eigenstates with positive eigenvalues from those with negative ones, 
$\cal{B}_S=\cal{B}_+ \cup \cal{B}_-$. Consequently, the Hilbert space can be expressed as a direct sum of the subspaces generated by $\cal{B}_+$ and $ \cal{B}_-$, respectively. That is $\cal{H}={\cal H}_+ \oplus {\cal H}_-$.   

Given two states of $\cal{H}$, $|x\rangle$ and $| y \rangle$, we define a new inner product:
\beqn
\langle x | y\rangle_{ S_K} = \langle x_+ | y_+ \rangle - \langle x_- | y_- \rangle,
\label{innerK}
\eeqn
where $|v_+ \rangle$ and $|v_- \rangle$ are the projections of $| v \rangle $ in the subspaces ${\cal H}_+$ and ${\cal H}_-$, respectively. 

The
new Hilbert space is the original vector space endowed with the inner product $\langle . | . \rangle_K$ defined in Eq.~\eqref{innerK}.

In order to simplify the problem, we construct a metric operator, $S_K$, such that $\langle . | . \rangle_{ S_K}=\langle S_K . | . \rangle$, as explained below.

We note that operator $S$ can be diagonalized
as $S= P D P^{-1}$, and split 
$D$ as 
$D= D_+ + D_-$, 
being $D_+$ and $D_-$ diagonal matrices with the positive and the negative eigenvalues, respectively. We define
\beqn
S_K = S_+ - S_-, ~~~S_\pm = P D_\pm P^{-1}.
\label{sK}
\eeqn
By construction $S_K$ is self-adjoint and positive definite, thus it can be written as $S_K= \Upsilon_K^\dag \Upsilon_K$, and
\beqn
\langle . | . \rangle_{ S_K}= \langle \Upsilon_K^\dag \Upsilon_K . | . \rangle \ .
\eeqn

\subsubsection{Exceptional points}
\label{sub213}

At the EPs, finite-dimensional systems are no longer diagonalizable, and to form a basis of the vector space, generalized eigenvalues should be included. Let us compute the Jordan form of $H^\dag$, {$H^\dag = 
 \ovl P J \ovl P^{\,-1}$}. Let $|\ovl \psi_k \rangle $ be the $k$-th column of $ {\ovl P}$ , and $| \ovl \nu_j \rangle $ the $j$-th column of $ {\ovl P^{\,-1}}$. The operator defined by
\beqn
    S = \sum_{\al \le \be}^{N} ~ \delta(\ovl{E}_\be-\wid{E}_\al^*) 
    \left( \zeta_\be \ketHC{\be} \langle \ovl \nu_\al| + \zeta_\be^* \ketHC{\al} \langle \ovl \nu_\be| \right) \nonn
    \label{seps}
\eeqn
can be seen to obey
the condition $H^\dag S = S H$. If $S$ in Eq.~\eqref{seps} is non-positive definite, we proceed as in the previous case  \ref{sub212}, and introduce an inner product after the construction of the metric operator in Krein space $S_J=\Upsilon_J^\dag \Upsilon_J$.

\subsection{Mean values and observables}
\label{sub22}

The metrics we have introduced for each of the possible situations can be written in a unified manner, as ${\cal S}= \gamma^\dag \gamma$. Then, the new inner product reads

\beqn
\langle . | . \rangle_{\cal S}=\langle . |{\cal S} . \rangle=\langle . | \gamma^\dag \gamma. \rangle=
\langle \gamma .| \gamma . \rangle \ .
\eeqn

As pointed out in Refs.~\cite{fring0,fring1}, to fix the metric we assume that the Hermitian operator $\hat{o}$  transforms as $\hat{O}= \gamma^{-1} \hat{o} \, \gamma$ \cite{Ram19}. So defined $\hat{O}^\dag S= S \hat O$. Consequently, the expression for the mean value of operator becomes 
\beqn
\langle . | \hat{O}| . \rangle_{\cal S}=\langle . | {\cal S }\hat{O} |. \rangle=\langle . |  \gamma^\dag \gamma \hat{O}| . \rangle
=\langle . | \gamma^\dag \hat{o} \,  \gamma | . \rangle =\langle \gamma  . |  \hat{o}   | \gamma. \rangle \ . \nonn
\eeqn

\subsection{Time evolution}
\label{time_evol}

We now turn to the description of the time evolution of an arbitrary initial state $|I(0)\rangle$.  
Expressed in a basis $\mathcal{A}_k=\{ |k\rangle \}$, it takes the form  
\begin{equation}
|I(0)\rangle = \sum_{k} c_k |k\rangle.
\label{eq-ini}
\end{equation}
Let us introduce the change to the basis $\mathcal{A}_H$ of eigenvectors (or generalized eigenvectors, at the EPs) of $H$, 
\begin{equation}
|I(0)\rangle = \sum_{\alpha} \tilde{c}_{\alpha} \ketH{\alpha} , 
\qquad 
\tilde{c}_{\alpha} = \sum_{k} Q_{k \alpha} c_k ,
\label{eq:init_Hbasis}
\end{equation}
where $Q$ denotes the transformation matrix relating the basis $\mathcal{A}_k$ with $\mathcal{A}_H$.  
We assume normalization of the initial state, $ \langle I(0)| I(0) \rangle_{\cal{S} }= 1$.  We choose the metric operator ${\cal S}$  according to the region in the space of parameters of the model. 

The time evolution of the 
state {of Eq.}~\eqref{eq:init_Hbasis} is given by
\begin{equation}
|I(t)\rangle ={\cal N}(t) \, \re^{-\uni Ht/ \hbar} |I\rangle 
= \sum_{\alpha} \tilde{c}_{\alpha}(t) \ketH{\alpha},
\label{eq:evo}
\end{equation}
where ${\cal{N}}(t)$ is the normalization constant at each time ~$t$, such that $\langle I(t)| I(t)  \rangle_{\cal{S}} = 1$.
Notice that if $H$ can be diagonalized, \ $\tilde{c}_{\alpha}(t) = \re^{-\uni E_{\alpha} t/\hbar} \, \tilde{c}_{\alpha}$. At exceptional points, we have to introduce a Jordan decomposition of $H$,
$\re^{-\uni Ht/ \hbar} = Q \, \re^{-\uni Jt/\hbar} Q^{-1}$, so that the evolution of $\tilde{c}_{\alpha}(t)$ depends explicitly on the structure of the corresponding Jordan blocks.  

Within this framework, the expectation value of an operator $\hat{o}$ at time $t$ is defined as  
\begin{equation}
\langle \hat{o}(t) \rangle = \langle I(t)| \, \hat{O} \, |I(t)\rangle_{\cal S} 
= \langle \gamma \, I(t)| \, \hat{o} \,  |\gamma \,I(t)\rangle 
\label{eq:mean_value}
\end{equation}
where the subscript $\cal S$ emphasizes that the scalar product is taken with respect to the modified inner product $\langle \cdot |\cdot \rangle_{\cal S}$ with 
${\cal{S}}= \gamma^\dagger\gamma$.

\section{Results and discussion}
\label{results}

As an application of the formalism proposed in the previous section, 
we 
study the following pseudo-Hermitian model Hamiltonian \cite{mandel}
\begin{equation}
\label{hamieq1}
H= r~ \re^{\uni \th}~\frac 12 ~(I+ \sg_z)+r ~ \re^{-\uni \th}~\frac 12 ~(I-\sg_z)+s~\sg_x,
\end{equation}
where $I$ is the identity matrix in two dimensions, and $\{ \sg_x,\sg_y,\sg_z \}$ are the Pauli matrices. We take $r$ and $s$ real, in units of energy, and $~ \th \in \mathbb R$.

The Hamiltonian in Eq.~\eqref{hamieq1} is invariant under the parity-time reversal transformation. In this case, the linear symmetry parity operator is given by $ \mathcal{P}= \sigma_{x}$ and the antilinear operator $\mathcal{T}$ is complex conjugation. This Hamiltonian was first introduced in Refs.~\cite{bender0,fring} in the context of time evolution of $\mcpt$-symmetry systems in the symmetry phase. More recently, experimental realizations of the system have been implemented \cite{Gao20} by using a single photon system to study its dynamical evolution. In the same line, in Ref.~\cite{gcanal}, the model has been implemented by a class of coupled resonant circuit chains.

We study the time evolution of a given initial state under the action of the Hamiltonian of Eq.~\eqref{hamieq1}. Particularly, we calculate the Survival Probability ($SP$) of the initial state $|I(0)\rangle$, that is, 
\beqn
SP(t) = |\langle I(0) | I(t) \rangle_S|^2.
\eeqn
Also, we will focus our attention on the modified Robertson uncertainty relation \cite{Rob29} for non-commuting observables \(A\) and \(B\). It can be expressed as
\begin{equation}
\Delta^2_S A \, \Delta^2_S B  \geq \frac{1}{4 } \big|\langle [A, B] \rangle_S \big|^2,
\label{eq:ModRobertson}
\end{equation}
with  $\Delta^2_S O= \langle {O}^2 \rangle_S -\langle {O} \rangle_S^2$.
The subscript \(S\) denotes that the expectation values and variances are computed with respect to the inner product \(\langle \cdot | \cdot \rangle_S\) in the Hilbert space.

{ {As an example with analytical solution, we consider the case of a system of one particle of spin $S=1/2$. In Appendix~\ref{app}, we present results for systems with larger numbers of particles.}}

The matrix representation of the Hamiltonian~\eqref{hamieq1} in the basis $\{|0\rangle= | 1/2,+1/2\rangle, \ |1\rangle= | 1/2,-1/2\rangle\} $ is given by
\begin{equation}
H=
\begin{pmatrix}
r \re^{\uni \theta} & s \\[6pt]
s & r \re^{-\uni \theta}
\end{pmatrix}.
\end{equation}

For $s^2-r^2 \sin^2\theta \neq 0$, it can be written as
\begin{eqnarray}
H & = &\wid{P} J \wid{P}^{-1}, \\
 \label{eqd}
J & = & 
\begin{pmatrix}
\rho - \lambda & 0 \\[6pt]
0 & \rho + \lambda 
\end{pmatrix},\\
\wid{P} & = & 
\begin{pmatrix}
-\lambda/s + \uni \eta & \lambda/s+ \uni \eta \\[6pt]
1 & 1 
\end{pmatrix}
\end{eqnarray}
with $\rho=r \cos\th$, $\lambda= \sqrt{s^2-r^2 \sin^2\th}$ and $\eta= (r/s) \sin\th$. 

From Eq.~\eqref{eqd}, it is clear that if $s^2-r^2 \sin^2\th > 0$ the spectrum is real, while for $s^2-r^2 \sin^2\th < 0$ we have two complex pair-conjugate eigenvalues.

Let us consider the case $s^2-r^2 \sin^2\th  =0$, that is $\sin\th= \pm s/r$ and $r \ge s$. In this case  
\begin{eqnarray}
H & = &\wid{P} J \wid{P}^{-1},\\
J & = & 
\begin{pmatrix}
\sqrt{r^2-s^2} & 1 \\[6pt]
0 & \sqrt{r^2-s^2} 
\end{pmatrix}, \label{eq:J} \\
\wid{P} & = & 
\begin{pmatrix}
 \uni  & 1/s \\[6pt]
1 & 0 
\end{pmatrix}.
\end{eqnarray}

\begin{figure}[h!]
     \centering
     \includegraphics[width=1.\linewidth]{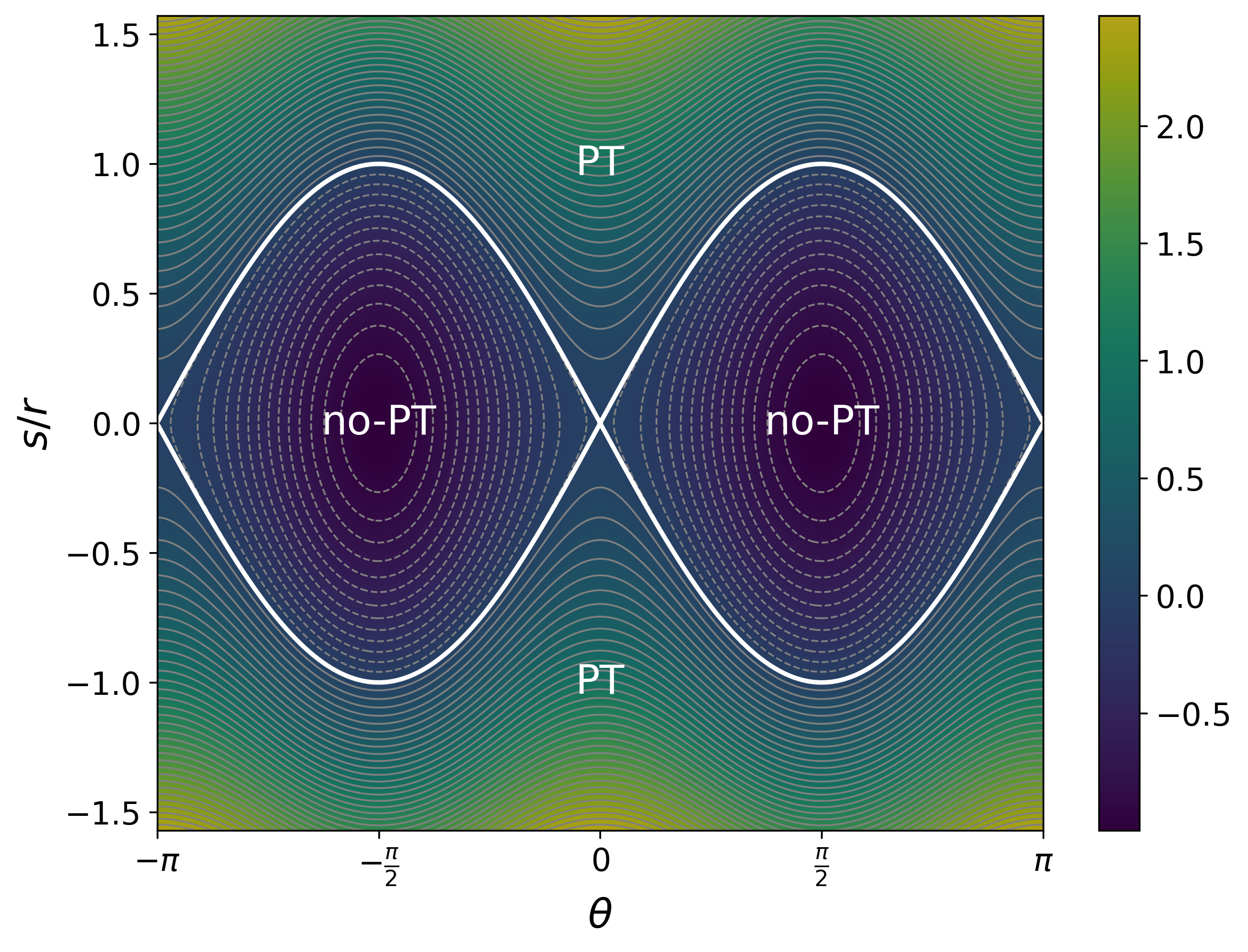}
     \caption{ 
    Dynamical phases of the model of Eq. ~\eqref{hamieq1} in terms of the values of $d=(s/r)^2-\sin^2\th$ in the plane $(\theta,s/r)$. The region for which $d>0$ corresponds to the  $\mcpt$-symmetry phase of the model, while the region with $d<0$ corresponds to the {{spontaneously}} broken symmetry phase. The white border between both regions corresponds to the localisation of the EPs, where $d=0$.}
     \label{fig:fases}
\end{figure}

In Figure~\ref{fig:fases} we plot $d=(s/r)^2-\sin^2\th$ in the plane $(\th,s/r)$. The region for which $d>0$ corresponds to the  $\mcpt$-symmetry phase of the model, while the region with $d<0$ corresponds to the {{spontaneously}} broken symmetry phase. The white border between the two regions corresponds to the localization of the EPs, where $d$ vanishes.

Let us analyze the time evolution of an initial state 
\beqn
|\steini_0\rangle = \frac{1}{\sqrt{1+p^2}} (|0 \rangle + p \re^{\uni \stang } |1 \rangle).
\label{ini}
\eeqn

\subsection{Exact symmetry phase}

In the $\mcpt$-symmetry phase the spectrum of the Hamiltonian $H$ in Eq.~\eqref{hamieq1} is real:
\beqn
E_\pm= \rho \pm \lambda,~~~ \lambda =s \sqrt{1-\eta^2}~\in {\mathbb R}.
\eeqn
We obtain
the similarity matrix by computing the eigenvectors of $H^\dag$ and calculate $S$ from Eq.~\eqref{spt}. It results
\beqn
S & = &
\Upsilon^\dag \Upsilon,
~~~\Upsilon=
\begin{pmatrix}
\eta_+  & \uni \eta_- \\[6pt]
-\uni \eta_-  & \eta_+
\end{pmatrix},
\eeqn
with $\eta_\pm=\frac 12 ( \sqrt{1 - \eta} \pm \sqrt{1 + \eta})$.

 We work in the Hilbert space endowed with the inner product $ \langle. | .\rangle_S $. Consequently, the initial state is normalized as 
$ \langle  \steini_0 | \steini_0 \rangle_S = \langle  \Upsilon \,\steini_0 | \Upsilon \,\steini_0 \rangle =1$. 

The results obtained can be summarized as follows

\begin{widetext}
\beqn
\langle \sg_x \rangle_S & = & \frac{\lambda}{s}  \frac {2 p \cos \stang}{1+p^2 + 2\,p\,\eta\,\sin \stang},\nonn 
\langle \sg_y \rangle_S & = & \frac{(1+p^2) \eta + 2 p \sin \stang}{1+p^2 + 2\,p\,\eta\,\sin \stang} \cos( 2\lambda t) - 
\frac{\lambda}{s} \frac{(1-p^2)}{1+p^2 + 2\,p\,\eta\,\sin \stang} \sin(2\lambda t),\nonn
\langle \sg_z \rangle_S & = & \frac{\lambda}{s} \frac{(1-p^2)}{1+p^2 + 2\,p\,\eta\,\sin \stang} \cos( 2\lambda t) + 
\frac{(1+p^2) \eta + 2 p \sin \stang}{1+p^2 + 2\,p\,\eta\,\sin \stang} \sin(2\lambda t),
\eeqn
\end{widetext}
and
\beqn 
\langle \sg_x^2 \rangle_S =\langle \sg_y^2 \rangle_S =
\langle \sg_z^2 \rangle_S=1.
\eeqn

\begin{figure*}[!htb]
     \centering
     \includegraphics[width=1.0\textwidth]{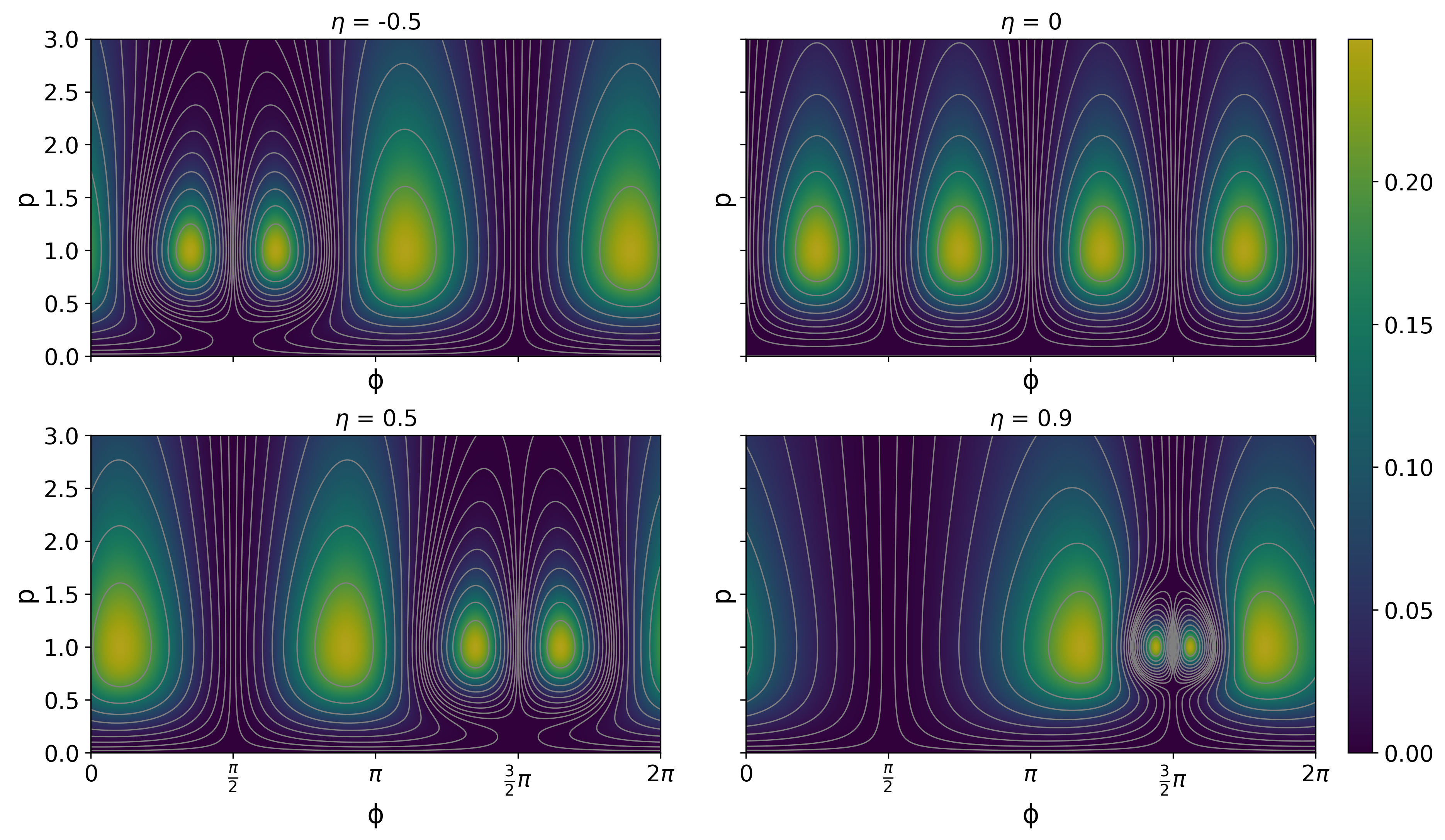}
     \caption{
     Contour plots of $UR(\sg_x,\sg_y)$, Eq.~\eqref{eq:URsgxsgy}, in the $(\stang,p)$-plane at $t=0$ for different values of $\eta$ within the $\PT$-symmetry region 
      ($\eta^2<1$).}
     \label{fig2}
\end{figure*}

In order to study 
the uncertainty relations for $A=\sg_x$ and $B=\sg_y, $ we compute the following quantity
\beqn
UR (\sg_x,\sg_y)= \Delta_S^2 \sg_x \, \Delta_S^2 \sg_y -|\langle  \sg_z \rangle_S|^2 \ge 0.
\label{eq:URsgxsgy}
\eeqn
In Figure~\ref{fig2} we present diverse
contour plots of $UR(\sg_x,\sg_y)$ in the $(\stang,p)$-plane at $t=0$ for different values of $\eta$ within the regime $\eta^2<1$. 

The survival probability at time $t$ in the unbroken symmetry phase, $\rm SP_{\PT}=|\langle \steini_0 | \steini (t) \rangle_S|^2$, is given by
%
\beqn
{\rm SP}_{\PT}= \cos^2(\lambda t) +
\frac {\lambda^2}{s^2} \frac{4 \, p^2  \cos^2\stang}{(1+p^2 + 2 \, p \, \eta \sin \stang)^2} \sin^2(\lambda t). \nonn
\eeqn

\subsection{{{Spontaneously}} broken symmetry phase}
\label{MRUR:noPT}

As discussed in the previous section, in this regime the operator $S$ in Eq.~\eqref{snpt} is not positive definite; therefore, the corresponding $S$-inner product becomes indefinite. To maintain a consistent physical interpretation, the analysis must be carried out using the spectral decomposition in a Krein space, as detailed in Sec.~\ref{sub212}. Following the spectral decomposition of the metric operator, we separate the contributions associated with the positive and negative eigenvalues to construct a positive-definite metric representation  $S_K$. It results
\beqn
S_K & = & 
\Upsilon_{K}^\dag \Upsilon_{ K},
~~~\Upsilon_{K}=
\begin{pmatrix}
\eta_+  & \uni \eta_- \\[6pt]
-\uni \eta_-  & \eta_+
\end{pmatrix},
\eeqn
where now
$\eta_\pm=\frac 12 ( \sqrt{|1 - \eta|} \pm \sqrt{|1 + \eta|})$. It should be noticed that in the { spontaneously} broken symmetry region we have $1-\eta^2<0$, that is either $1-\eta <0 ~\land~ 1+\eta >0$ \ or \ $1-\eta >0 ~\land~ 1+\eta <0$. Both cases are considered defining $\eta_\pm$. Also, for this region we have $\lambda= s \sqrt{\eta^2-1}$.

We have computed mean values of observables and the time evolution of the initial state of Eq.~\eqref{ini} in the Hilbert space endowed with the inner product $\langle . | . \rangle_{S_K}$. They read

\begin{widetext}
\beqn
\langle \sg_x \rangle_{S_{K}} & = & {\cal N}(t)^2  \frac{\lambda}{s} 2\, p  \cos \stang, \nonn
\langle \sg_y \rangle_{S_{K}} & = & {\cal N}(t)^2  (1+p^2 + 2 \, p \,\eta \sin\stang), \nonn
\langle \sg_z \rangle_{S_{K}} & = & {\cal N}(t)^2 \left( \frac {\lambda}{s} (1-p^2) \cosh(2 \lambda t) +
 [(1+p^2) \eta  + 2 \, p \,\sin\stang] \sinh( 2\lambda t) \right), 
\eeqn
with
\beqn
{\cal N}(t)^2 = \left( [(1+p^2) \eta  + 2 \, p \sin \stang] \cosh( 2 \lambda t) +\frac{\lambda}{s} (1-p^2) \sinh(2 \lambda t)\right)^{-1},
\eeqn 
\end{widetext}
and
\beqn 
\langle \sg_x^2 \rangle_{ S_{K}} =\langle \sg_y^2 \rangle_{ S_{K}} =\langle \sg_z^2 \rangle_{S_{K}}=1.
\eeqn

\begin{figure*}[!htb]
     \centering    
     \includegraphics[width=1.0\textwidth]{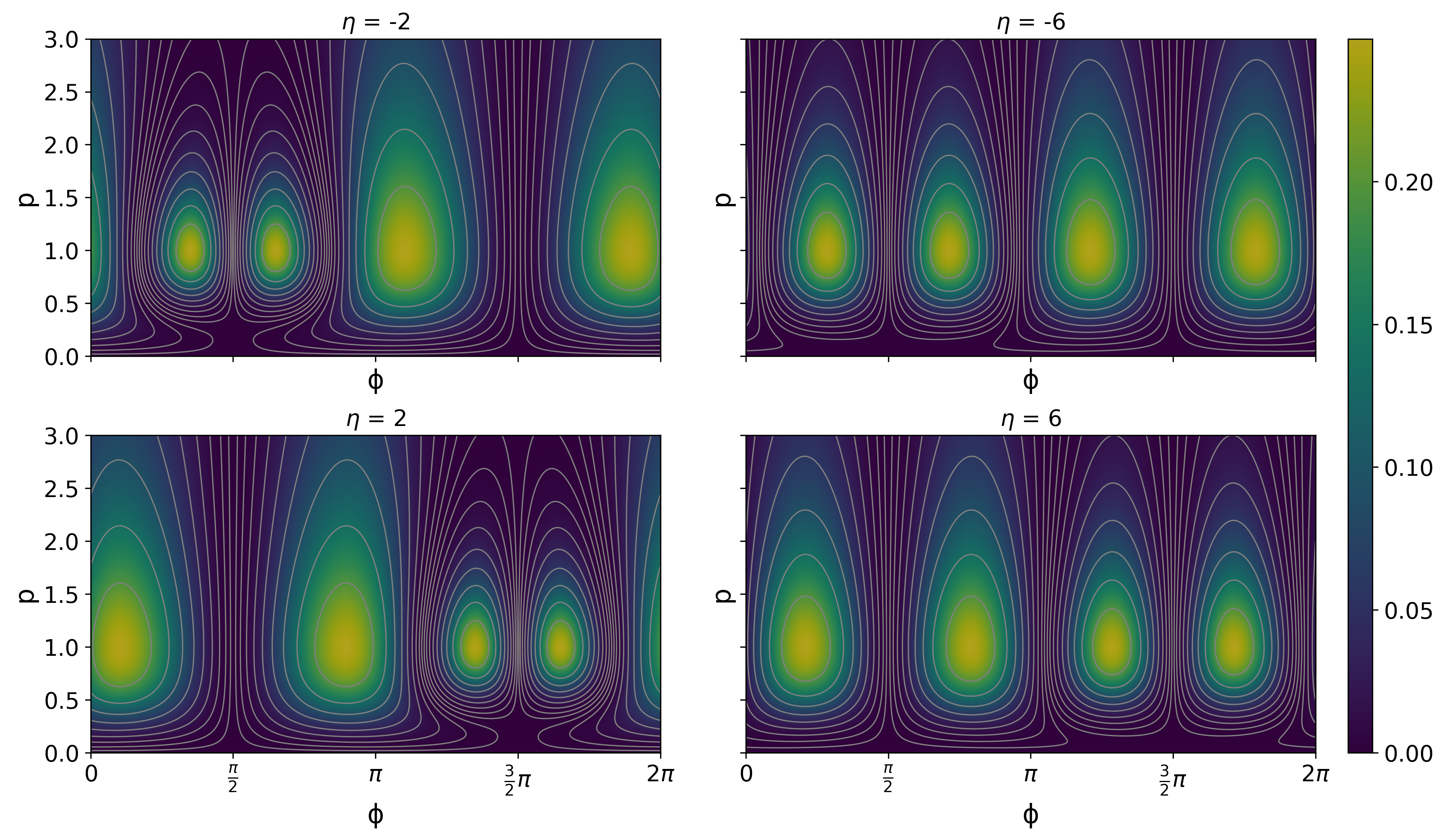}
     \caption{Contour plots of $UR(\sg_x,\sg_y)$, Eq.~\eqref{eq:URsxsynoPT}, in the $(\stang,p)$-plane at $t=0$ for different values of $\eta$ within the {{spontaneously}} broken symmetry region ($\eta^2>1$).}
     \label{fig3}
\end{figure*}

The uncertainty difference
for $A=\sg_x$ and $B=\sg_y$ is now computed as: 
\beqn
UR (\sg_x,\sg_y)= \Delta^2_{S_{K}}´ \sg_x \, \Delta_{  S_{K}}^2 \sg_y -|\langle  \sg_z \rangle_{ S_{K}}|^2 \ge 0.
\label{eq:URsxsynoPT}
\eeqn
In Figure~\ref{fig3}, we present various contour plots of $UR(\sg_x,\sg_y)$ in the $(\stang,p)$-plane at $t=0$ for different values of $\eta$ within the regime $\eta^2>1$.

The survival probability at time $t$ in the {{spontaneously}} broken symmetry phase, ${\rm SP_{\sim \PT}}=| \langle \steini_0 | \steini (t) \rangle_{S_{K}}|^2 $, is given by

\begin{widetext}
\beqn
{\rm SP_{\sim\PT}}= {\cal{N}}(0)^2 {\cal{N}}(t)^2 \left([(1+p^2)\eta+2 \, p \sin\stang]\cosh(\lambda t)+(1-p^2) \frac{\lambda}{s} \sinh(\lambda t) \right)^2
\label{eq:SPnonPT}
\eeqn
\end{widetext}

In the limit of large values of $t$, the survival probability tends to the value:
\beqn
{\rm SP_{\sim \PT}} \ \rightarrow \  
\frac 12 \left( 1+
\frac{(1-p^2) \sqrt{\eta^2-1}}
{(1+p^2) \eta +2 \, p \sin\stang}
\right). 
\eeqn
It has maximum values for 
$(\stang,~ p)$ equal to \ $(\pi/2,~\eta \pm\sqrt{\eta^2-1})$ \ and \  
$(-\pi/2,~\eta \pm\sqrt{\eta^2-1})$.

It is worth noting that, in the spontaneously broken symmetry phase, the eigenstates of $H$ do not satisfy an orthogonality condition with respect to the $S_K$-inner product, that is, $\langle \steH_n | \steH_m \rangle_{S_K} \ne \delta_{mn}$. This behavior has been observed by adopting the conventional inner product in Ref.~\cite{joglekar2019}. 

\begin{figure}[h!]
\centering
\includegraphics[width=0.5\textwidth]{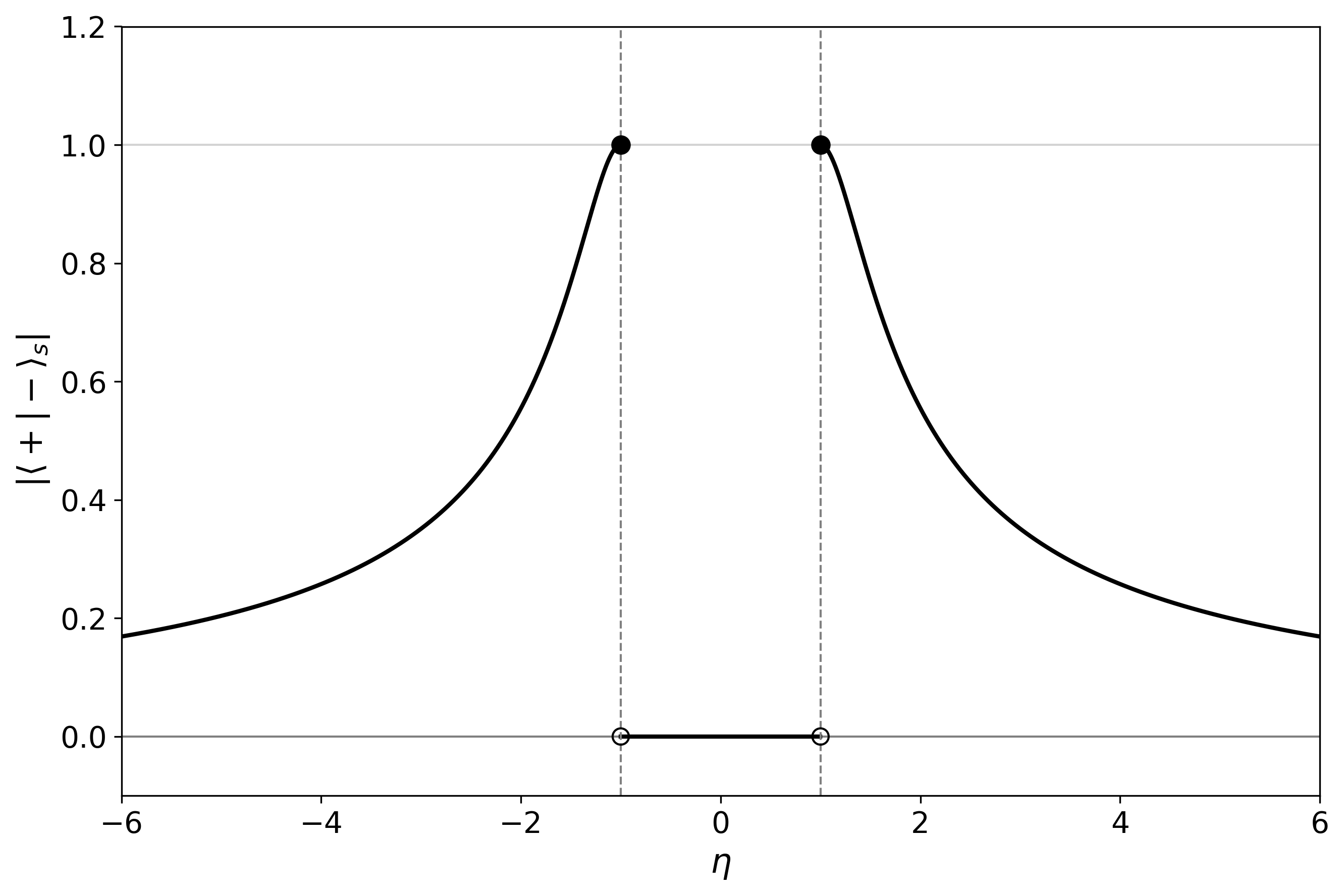}
\caption{Overlap $|\langle + |- \rangle _S|$ between the states with energies $E_+$ and $E_-$, as a function of $\eta$, Eq.~(\ref{eq:over}).}
\label{fig4}
\end{figure}

In Figure~\ref{fig4} we show 
the overlap $|\langle v_+ |v_- \rangle _S|$, given by Eq.~(\ref{eq:over}), between the states with energies $E_+=\rho+\uni \lambda$ and 
$E_-=\rho-\uni \lambda$, as a function of $\eta$, . 
It can be seen that indeed when $|\eta| >1$, this overlap takes values in the interval $(0,1)$. For completeness the figure also exhibits the results at the exceptional points ($|\eta|=1$) and in the symmetry region ($|\eta|<1$).
It can be seen that at the EPs  
both eigenfunctions are coalescent, and that in the $\PT$-symmetry phase the states are orthogonal with respect to the inner product induced by the metric of Eq.~\eqref{spt}.

\beqn
 |\langle + |- \rangle _S|=  \begin{cases} 0 & \text{if } |\eta| < 1 \\ 1 & \text{if } |\eta| =1 \\  \frac{|\eta|}{\sqrt{1-\eta^2+\eta^4}} & \text{if } |\eta| > 1 \end{cases}
 \label{eq:over}
\eeqn

\subsection{Exceptional points}
\label{UR:EP}

At the exceptional points, the Hamiltonian is no longer diagonalizable and must instead be expressed in its Jordan form. The algebraic structure of the system at the EP can be constructed through its Jordan decomposition, as detailed above
in Sec.~\ref{sub213}.
According to Eq.~\eqref{seps}, in the basis of generalized eigenvectors the operator $S_J$ takes the form
\beqn
S_{J} =
\begin{pmatrix}
0 & b/s\\
b/s & 0
\end{pmatrix},
\label{seps1}
\eeqn
where $b$ is chosen so that $b/s>0$. It can be seen that
$S_J$ 
satisfies $\HC S_J= S_J H$.

The metric operator $S_{ J}$ can be written as
\beqn
S_{J} & = & \Upsilon_{ J}^\dag \Upsilon_{ J}, \nonn
\Upsilon_{ J} & = &
\begin{pmatrix}
\sqrt{b/s} & 0 \\
0 & \sqrt{b/s}
\end{pmatrix}.
\eeqn
We choose, without loss of generality, $b = 1$ and $s > 0$ (or $b =- 1$ and $s < 0$).


Normalizing $\braketbar{ \Upsilon_J\,\steini}{ \Upsilon_{J}\,\steini}=1$ and using the $S_{J}$-inner product, we obtain

\begin{widetext}
\beqn
\langle \sg_x \rangle_{S_J} & = & {\cal N}(t)^2 2 \, p \cos\stang, \nonn
\langle \sg_y \rangle_{S_J} & = & {\cal N}(t)^2 
\left (2 \, p \sin \stang - 2 (1-p^2) s t - 2 (s t)^2 ( \eta \, (1+p^2)+ 2 \, p \sin\stang) 
\right), \nonn
\langle \sg_z \rangle_{S_J} & = & {\cal N}(t)^2  
\left (  1-p^2 + 2 s t(\eta \, (1+p^2) + 2 \, p \sin\stang)
\right), \nonn
\eeqn
with
\beqn
{\cal N}(t)^2 = \left(  
1+ p^2 + 2 \, \eta \, (1-p^2) s t + 2 (s t)^2 (1+p^2 +2\, \eta \, p \sin\stang)
\right)^{-1} \nonn
\eeqn
\end{widetext}

and 
\beqn 
\langle \sg_x^2 \rangle_{S_J} =\langle \sg_y^2 \rangle_{ S_J} =\langle \sg_z^2 \rangle_{ S_J}=1.
\eeqn

\begin{figure}[h!]
     \centering
     \includegraphics[width=1.0\linewidth]{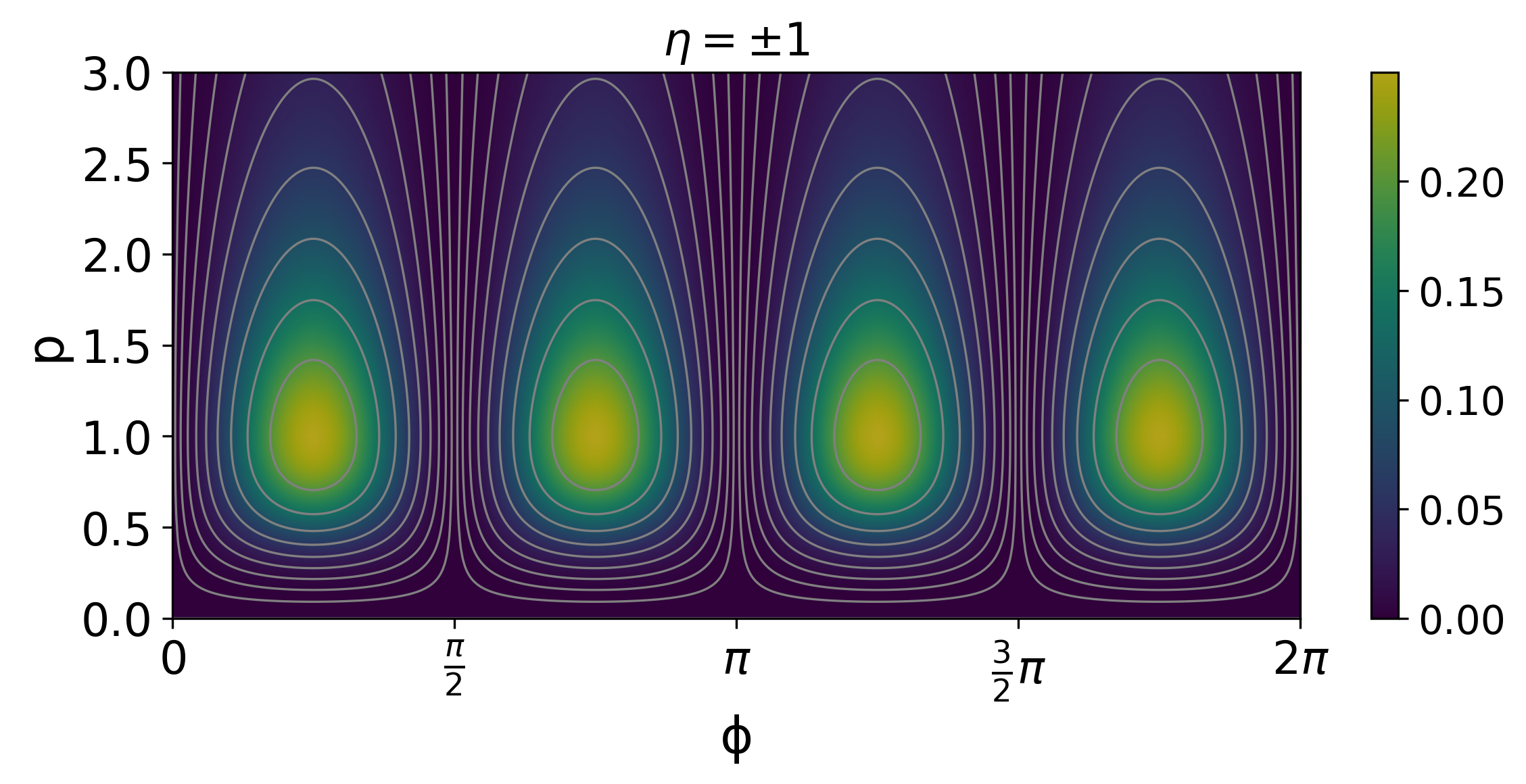}
     \caption{Contour plots of $UR(\sg_x,\sg_y)$, Eq.~\eqref{eq:URsxsyEP},  in the $(\stang,p)$-plane at $t=0$ at the exceptional points $(\eta=\pm 1)$.}
     \label{fig5}
\end{figure}

The uncertainty difference for $A=\sg_x$ and $B=\sg_y$ in this case is given by 
\beqn
UR (\sg_x,\sg_y)= \Delta^2_{S_{J}} \sg_x \, \Delta_{ S_{J}}^2 \sg_y -|\langle  \sg_z \rangle_{ S_{J}}|^2 \ge 0.
\label{eq:URsxsyEP}
\eeqn
In Figure~\ref{fig5}, we present various contour plots of $UR(\sg_x,\sg_y)$ in the $(\stang,p)$-plane at $t=0$ for $\eta=\pm 1$. 

\begin{widetext}
The survival probability at time $t$ at the exceptional points, ${\rm SP_{EP}}=| \langle \steini_0 | \steini (t) \rangle_{S_{J}}|^2$, reads
\beqn
{\rm SP}_{EP} & = & {\cal N}(0)^2 {\cal N}(t)^2 
\left ((1+p^2)^{ 2} +2 \, \eta \, (1-p^4) s t + (s t)^2 [1+p^4-2  \, p^2 \cos(2 \stang)] 
\right) . 
\label{eq:SPEP}
\eeqn
\end{widetext}
In the limit of large values of $t$, the survival probability approaches the value 
\beqn
{\rm SP}_{EP}  \rightarrow \frac{1+p^2- 2 \, \eta \, p \sin\stang}{2(1+p^2)},
\eeqn

It reaches its maximum when $\sin \stang=-\eta \, \text{sgn}(p)$ that is, for $\stang=-\pi/2$ if $\eta \, p >0$ and $\stang=\pi/2$ if $\eta \, p <0$ .

\subsection{Temporal behavior in the three regimes}

\begin{figure}[h!]
     \centering
     \includegraphics[width=1.0\linewidth]{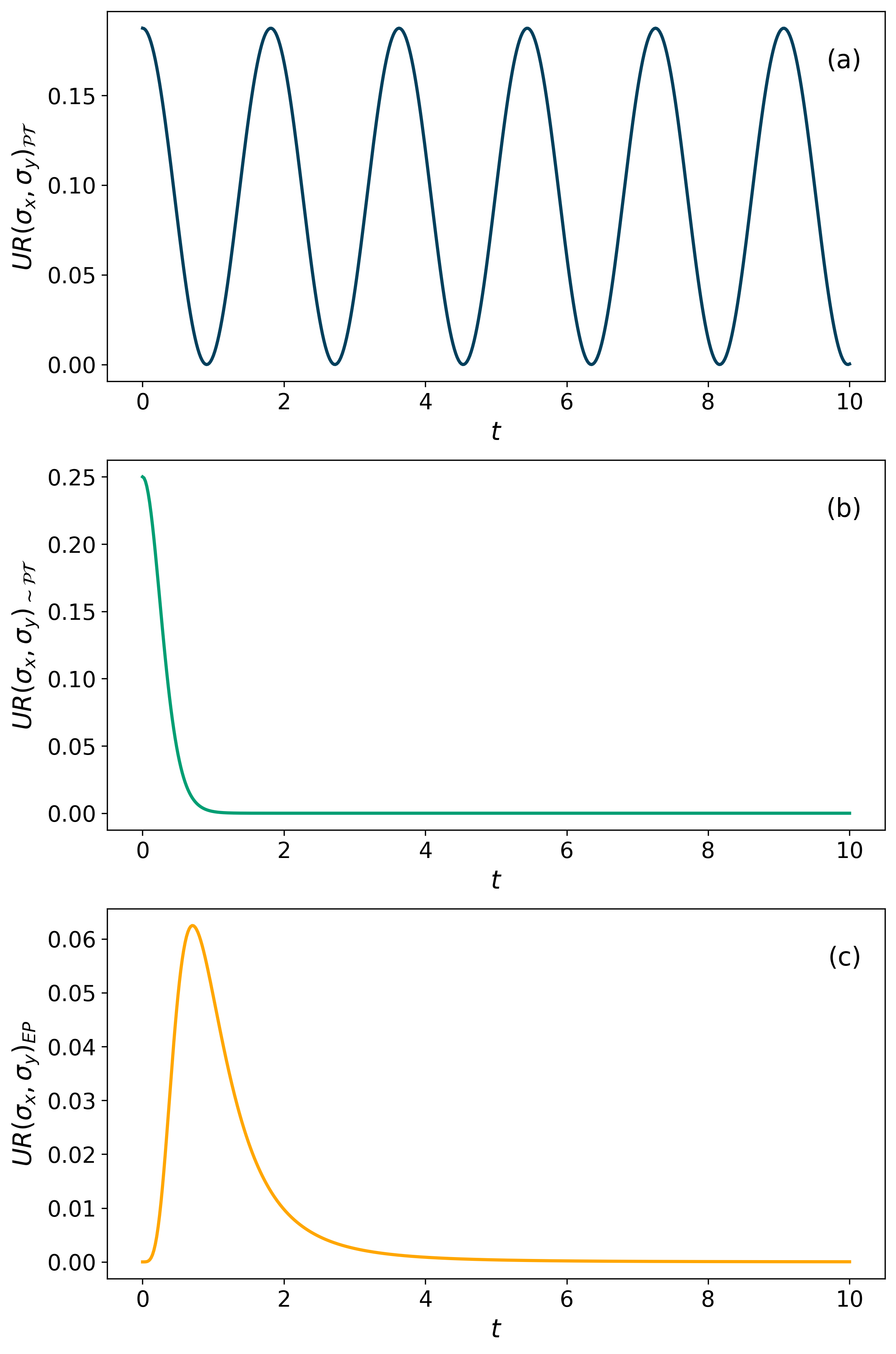}
     \caption{$ UR(\sg_x,\sg_y)$ as a function of time. Panel~(a) shows the results obtained for $\eta=1/2, s=1$ in the $\PT$-symmetry phase. In panel~(b), the results depicted correspond to $\eta=\sqrt{2}, s=1 $, in the { {spontaneously broken }}symmetry phase. In panel~(c), we show the results obtained for one of the EPs, with $\eta =1, s=1$. The initial state has been fixed to $(\stang,p)= (\pi,1)$.}
     \label{fig6}
\end{figure}

In this subsection we present the evolution with time of the relevant quantities defined above, for the different regimes.
In Figure~\ref{fig6}, we show the behavior of $UR(\sg_x,\sg_y)$ as a function of time. The initial state parameters 
have been fixed at $(\stang, p)= (\pi,1)$. In panel~(a) the results depicted correspond to the $\PT$-symmetry phase,  with $\eta=1/2,~s=1$; panel~(b) displays the results for the { {spontaneously broken }}-symmetry phase, with $\eta=\sqrt{2},~s=1 $; and panel~(c) presents results for the exceptional points with $\eta=1, ~s=1$. As expected, in the $\PT$-symmetry phase the behavior is oscillatory, while in the { {spontaneously broken }}-symmetry and at the EP, as the initial state evolves in time it behaves as a steady state with minimum uncertainty.

Up to here, the operators $A$ and $B$ of Eq.~\eqref{eq:ModRobertson} have been arbitrarily chosen. As pointed out in Ref.~\cite{KU93}, for a system of spins it is important to establish the uncertainty relations in the plane normal to the mean value of the spin at each instant in time. That is 
\beqn
\Delta^2 \sg_{x'} \Delta^2 \sg_{y'} \ge  |\langle \sg_{z'} \rangle |^2,
\eeqn
with $z'$ in the direction of the mean value of $\vec{\sg}$, $x'$ the direction of minimum uncertainty in the plane perpendicular to $\langle \vec{\sg}\rangle$, and $y'$ perpendicular to both directions.

The mean value of $\vec{\sg}$ at each instant of time is given by
\beqn
\langle \vec{\sg} \rangle_S = (\langle \sg_x \rangle_S,\langle \sg_y \rangle_S, \langle \sg_z \rangle_S).
\eeqn
It is straightforward to prove that $|\langle \vec\sg \rangle_S|=1$ at  all time $t$ and 
for both dynamical phases and the EPs.

\begin{figure}[h!]
     \centering
     \includegraphics[width=1.0\linewidth]{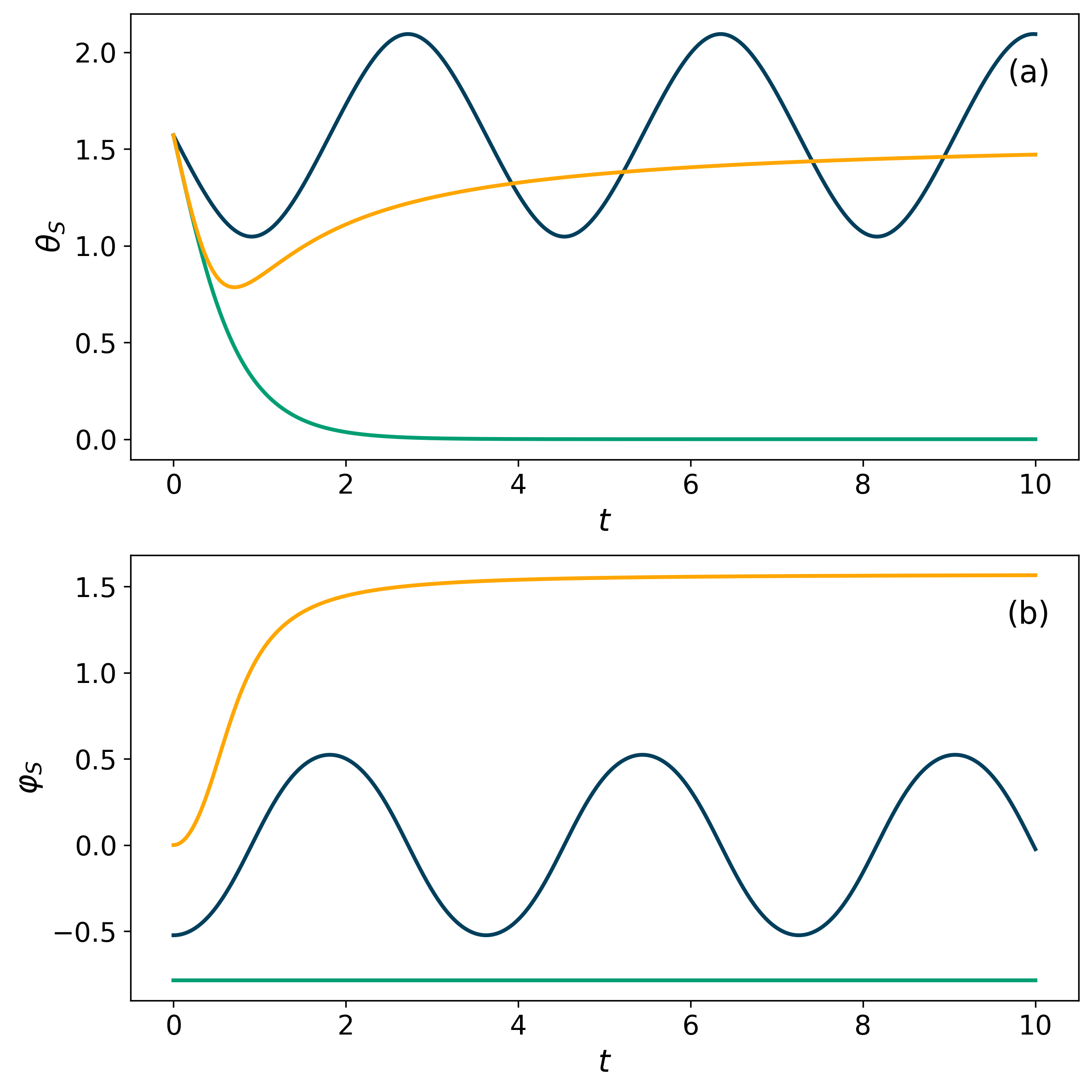}
\caption{Panel (a) shows the angle $\theta_S$ and panel (b) shows the angle $\varphi_S$, both defined in Eq.~\eqref{angulos}, as a function of time. The blue curves correspond to the $\PT$-symmetry phase, obtained for $\eta=1/2, s=1$. The green curves depict the { {spontaneously broken }} symmetry phase results, taking $\eta=\sqrt{2}, s=1 $. The yellow curves  show the results obtained for one of the EPs, with $\eta =1, s=1$. The initial state has been fixed to $(\stang,p)=(\pi,1)$. }
     \label{fig7}
\end{figure}

The direction $z'$ is determined, at each instant of time, by two angles, $(\th_s, \varphi_s)$, given in terms of the mean values of the components of $\vec{\sg}$:
\beqn
\th_s &=& \arccos \, ( \langle \sg_z \rangle_S), \nonn
\varphi_s & = & \arctan \left ( \frac { \langle  \sg_y \rangle_S }{\langle \sg_x \rangle_S }\right ).
\label{angulos}
\eeqn
In Figure~\ref{fig7} we plot the angles $\theta_S$ and $\varphi_S$ as a function of time, for different sets of parameters.

After some algebra it can be proved that, in the system given by directions $(x',y',z')$, the uncertainty relation has the minimum possible value, saturating the inequality. Indeed 
\beqn
\Delta_S^2 \sg_{x'} \, \Delta_S^2 \sg_{y'}=1,
\eeqn
with 
\beqn
\Delta_S^2 \sg_{x'}=\Delta_S^2 \sg_{y'}=1.
\eeqn
This means that, in the plane perpendicular to the mean value of spin, the state of Eq.~\eqref{ini} evolves in time as a coherent spin state. This result does not depend on the region in the space of parameters of the model. { In Appendix \ref{app}, we discuss results obtained for systems \MR{with larger numbers} of particles.}

\begin{figure}[h!]
     \centering
     \includegraphics[width=1.0\linewidth]{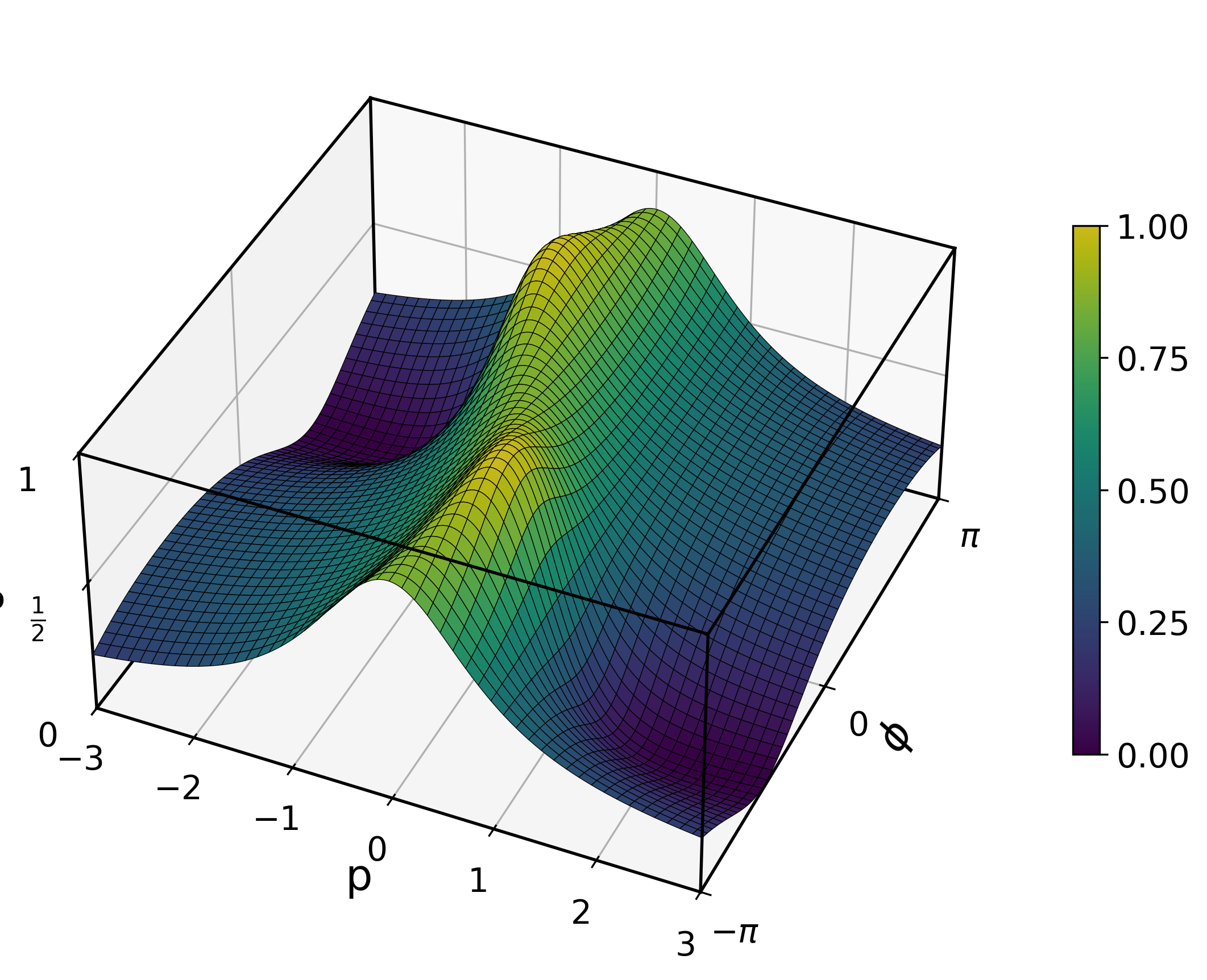}
     \caption{Stationary value of the  survival probability in the {{spontaneously}} broken symmetry phase, $SP_{\sim PT}$, as a function of the parameters $(\stang,p)$ of the initial state of Eq.~\eqref{ini} and for $\eta=\sqrt{2}$.}
     \label{fig8}
\end{figure}

In order to complete our analysis, in Figure~\ref{fig8} we present the value of the survival probability at the limit of large values of time as a function of the parameters of the initial state of Eq.~\eqref{ini},  $\stang$ and $p$, in the {{spontaneously}} broken symmetry phase for $\eta=\sqrt{2}$. It can be observed that the survival probability takes its maximum value $SP_{\sim PT}=1$ when  
$(\stang,p)=\pm (\pi/2, \sqrt{2}-1)$; 
for these values of the parameters,  
the initial state is preserved under the time evolution induced by the Hamiltonian of Eq.~\eqref{hamieq1}. 

\begin{figure}[h!]
     \centering
     \includegraphics[width=1.0\linewidth]{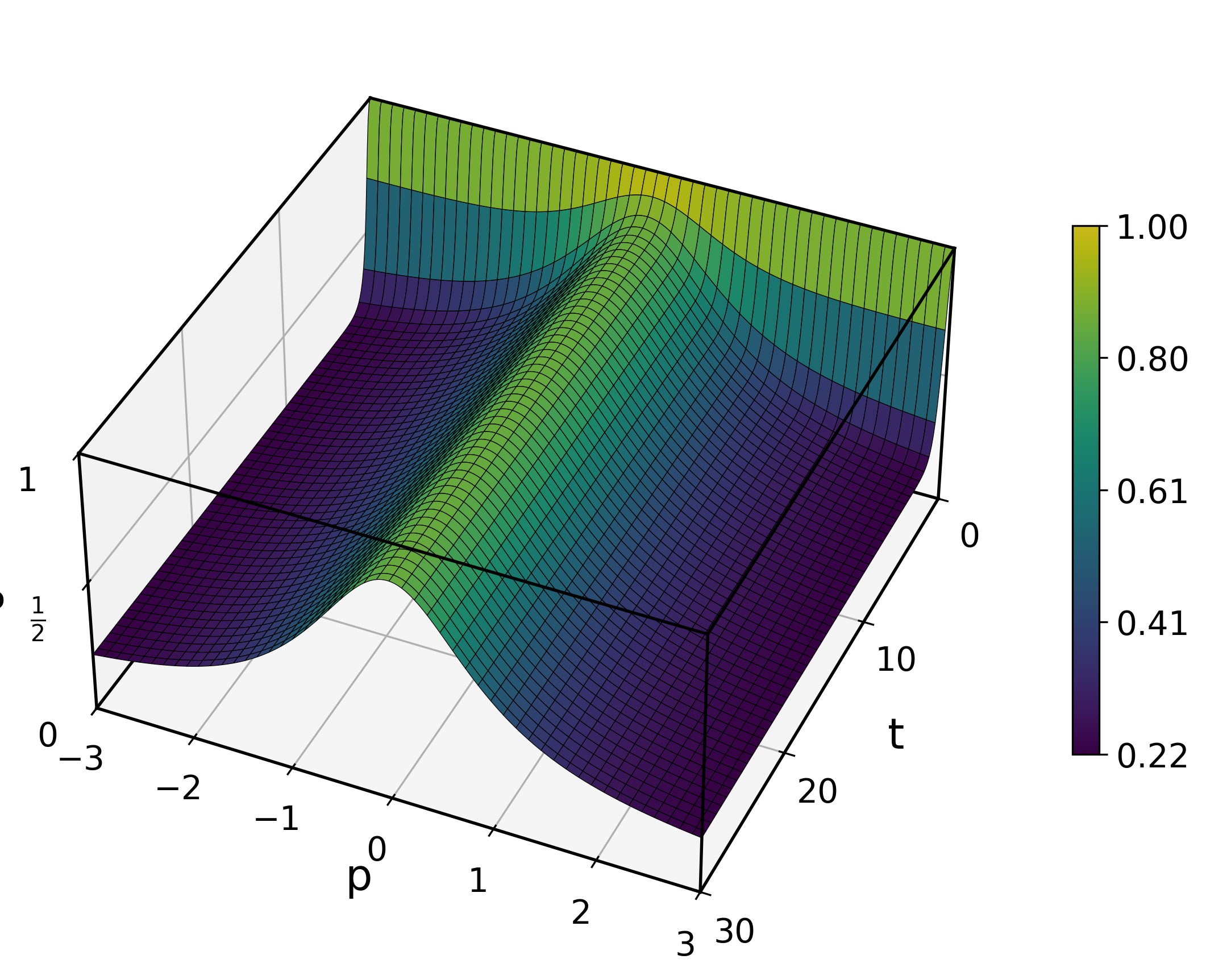}
     \caption{Survival probability in the {{spontaneously}} broken symmetry phase, ${\rm SP}_{\sim PT}$ of Eq.~\eqref{eq:SPnonPT}, as a function of $p$ and $t$, for $\eta=\sqrt{2}$, $\stang=-\pi/2$, and $s=1$.}
     \label{fig9}
\end{figure}

The behavior of ${\rm SP}_{\sim PT}$, Eq.~\eqref{eq:SPnonPT}, in terms of $p$ and $t$ is depicted in Figure~\ref{fig9}, taking $\eta=\sqrt{2}$, $\stang=-\pi/2$, and $s=1$.

\begin{figure}[h!]
\centering
\includegraphics[width=1.0\linewidth]{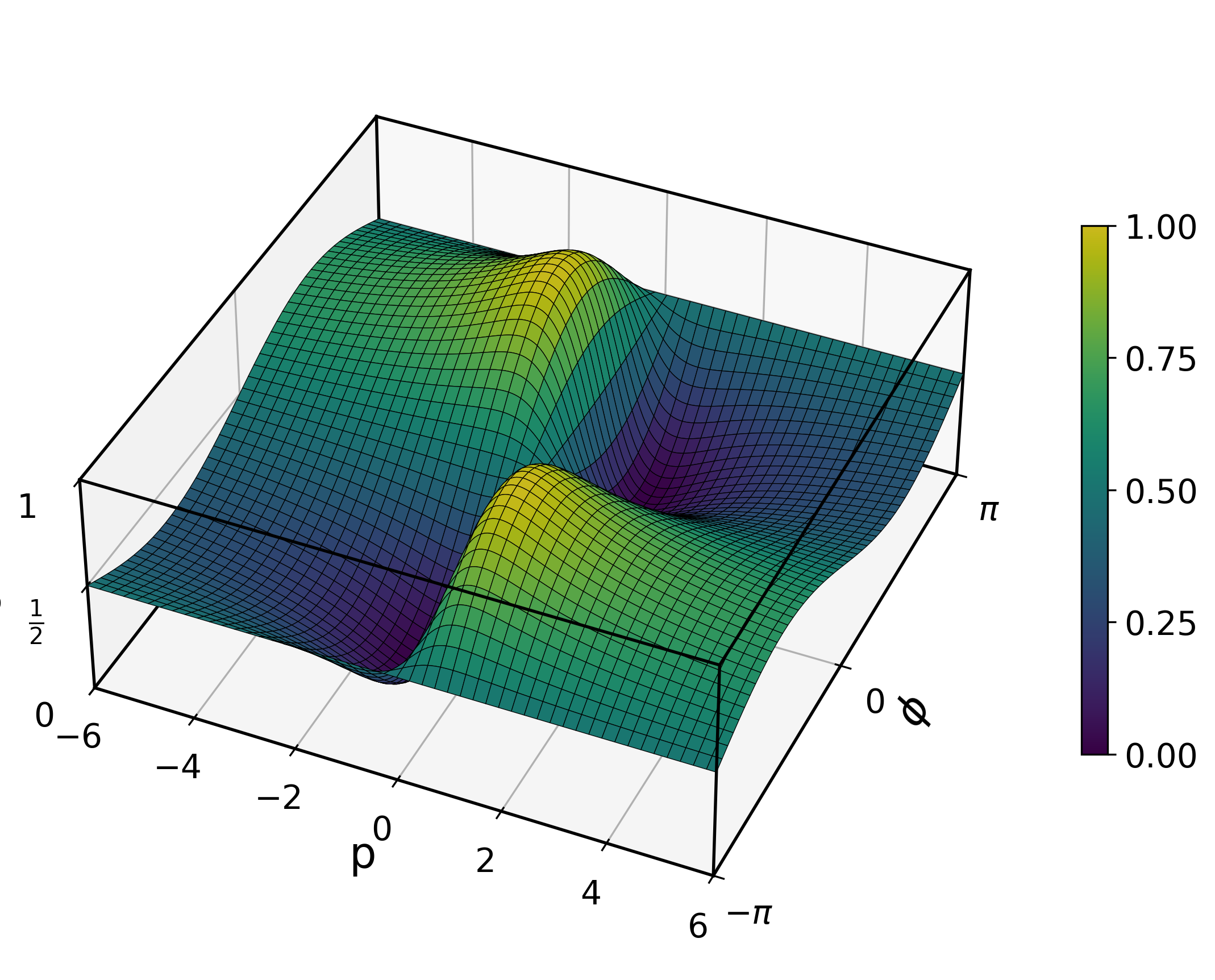}
\caption{Stationary value of the survival probability at an exceptional point, ${\rm SP}_{EP}$, as a function of the parameters $(\stang,p)$ of the initial state of Eq.~\eqref{ini} and for $\eta=1$, \st{$s=1$}.}
\label{fig10}
\end{figure}

In Figure~\ref{fig10} the value of the survival probability at the limit of large values of time is depicted as a function of the parameters of the initial state of Eq.~\eqref{ini}, $\stang$ and $p$, at the exceptional point with $\eta=1$. It can be observed that its maximum value is $\rm SP_{EP}=1$, occuring at $(\stang, p)=(-\pi/2,1)$ and $(\pi/2, -1)$; for these values of the parameters, the initial state is preserved under the time evolution induced by the Hamiltonian of Eq.~\eqref{hamieq1}.

\begin{figure}[h!]
     \centering
     \includegraphics[width=1.0\linewidth]{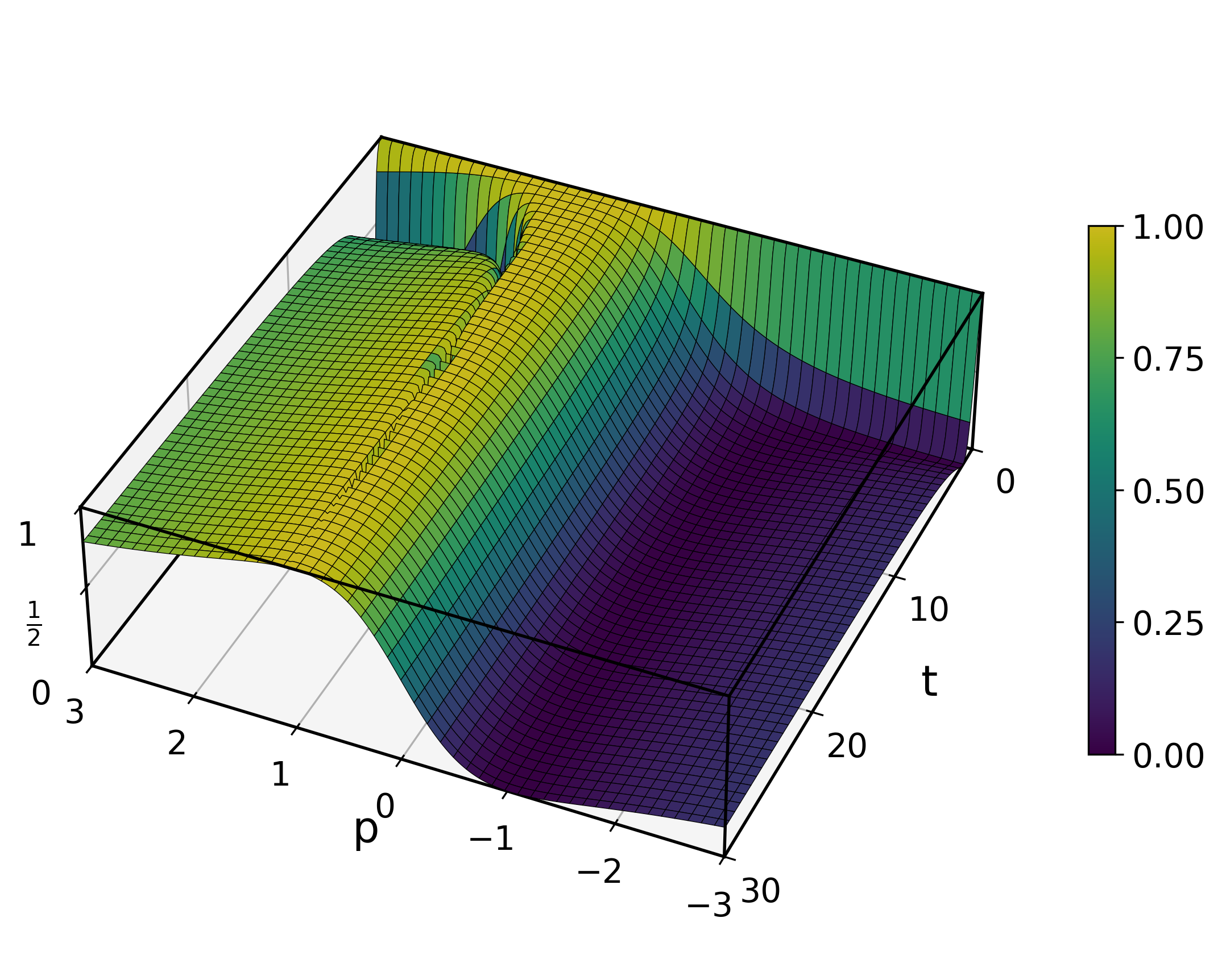}
     \caption{Survival probability at the exceptional point, ${\rm SP}_{EP}$ of Eq.~\eqref{eq:SPEP}, as a function of   $p$ and $t$, for $\eta=1$, $\stang=-\pi/2$, and $s=1$.}
     \label{fig11}
\end{figure}

The behavior of ${\rm SP}_{EP}$, Eq.~\eqref{eq:SPEP}, as a function of $p$ and $t$ is depicted in Figure~\ref{fig11}, for $\eta=1$, $\stang=-\pi/2$, and $s=1$.

\subsection{Comparison with the Lindblad formalism} 

To demonstrate that our approach is consistent with the expected physical behavior of the model, we compare our results in the {{spontaneously}} broken symmetry phase with those obtained using the Lindblad formalism. 

We have analyzed the results obtained from the Lindblad equation 
\beqn
\frac{d\rho}{dt} = -\frac{\uni}{\hbar} [h, \rho] + \sum_{k} \gamma_k \left( L_k \rho L_k^\dagger - \frac{1}{2} \left\{ L_k^\dagger L_k, \rho \right\} \right),
\label{lind}
\eeqn
with $h$ given by
\begin{equation}
\label{hamieq0}
h= r\,\cos\th~I+s\,\sg_x.
\end{equation}
We have used as collapse operators $L\pm= (\sg_x \pm \uni\, \sg_y)/2$. Finally, $\gamma_k= \sqrt{|r \sin \th|}$. 

\begin{figure}[h!]
\centering
\includegraphics[width=1.0\linewidth]{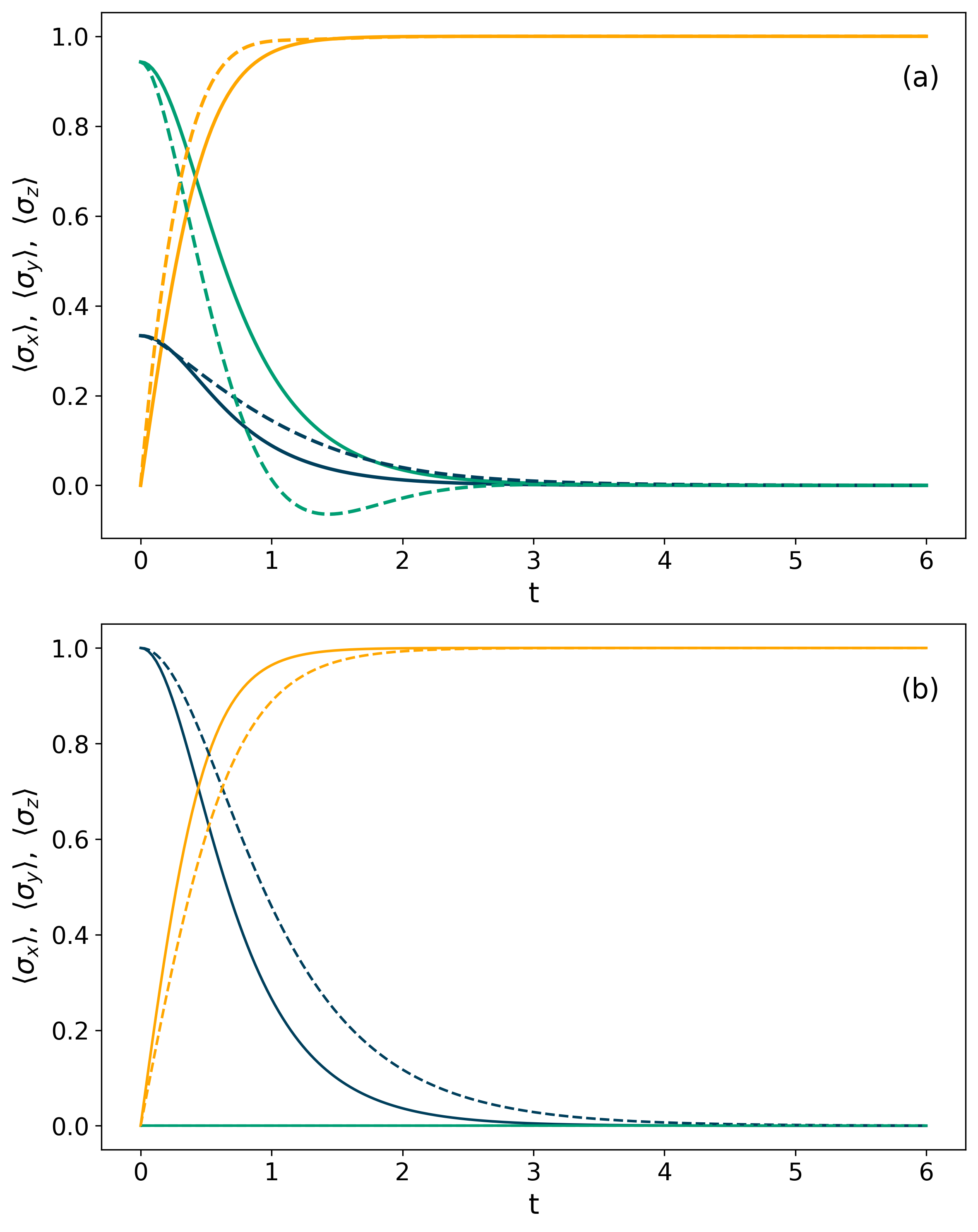}
\caption{Behavior of the mean values of the components of spin, $\langle \sg_j \rangle_{S_K}  \ (j=x,y,z)$, as a function of time ( blue, green and yellow curves correspond to $\la \sg_x \ra$, $\la \sg_y \ra$ and $\la \sg_z \ra$, respectively). Solid lines show  the results obtained with the formalism presented in Section~\ref{formalismo}, while dotted lines correspond to results obtained from the Lindblad equation. The parameters of the initial state are taken as $p=1$, $\stang=\pi/4$ in panel~(a), and $p=1$, $\stang=-\pi/4$ in panel~(b). The parameters that model the  interaction have been fixed to $r=\sqrt{2}$, $s=1$ and $\th=\pi/2$.
}
\label{fig12}
\end{figure}

In Figure~\ref{fig12} we plot the behavior of the mean values of the components of spin, $\langle \sg_j \rangle_{S_K} \ (j=x,y,z)$, as a function of the time. Solid lines correspond to the results obtained with our formalism, while dotted-line show the  results obtained from the Lindblad equation. The parameters of the initial state correspond to values of $\stang=\pm \pi/4$ and $p=1$, in panels~(a) and~(b) respectively; while the parameters that model the 
interaction have been fixed to $r=\sqrt{2}$, $s=1$ and $\th=\pi/2$. 
From the figure it can be observed that, as the initial 
state evolves in time, both formalisms predict the same steady-state behavior.
The differences at early times are due to the inclusion of quantum jumps in Eq.~\eqref{lind}. The formalism presented in this work assumes a postselection mechanism for the time evolution of the system \cite{joglekar2019,nori}.

{
\MR{For the sake of completeness, beyond the two-level case we have addressed  a system consisting of $N \ (N>2)$ levels governed by the Hamiltonian~(\ref{hamieq1}). These results are shown and discussed in Appendix~\ref{app}.}
}

\section{Concluding remarks}
\label{conclusions}

We have presented a consistent formulation of uncertainty relations for quantum observables evolving under non-Hermitian Hamiltonians, emphasizing the central role played by metric operators in defining physically meaningful quantities. By explicitly constructing metrics adapted to each dynamical regime —namely the unbroken-symmetry phase, the {{spontaneously}} broken-symmetry phase, and exceptional points— we have shown that expectation values, variances, and time evolution can be defined in a unified manner across all spectral scenarios.

Within this framework, we derived a generalized Heisenberg-Robertson uncertainty inequality expressed in terms of suitable metric-modified inner products. This construction ensures the validity of the uncertainty relation even in regimes where the Hamiltonian spectrum contains complex eigenvalues or the eigenvalues are coalescent (see Eqs.~(\ref{eq:URsgxsgy}),~(\ref{eq:URsxsynoPT}), and (\ref{eq:URsxsyEP})). Our results demonstrate that the inclusion of appropriate metric structures is not optional but rather essential for maintaining a consistent physical interpretation in the context of non-Hermitian quantum dynamics.

As an explicit application, we analyzed a $\PT$-symmetric two-level system and studied the behavior of uncertainty relations across different spectral phases. In the unbroken-symmetry phase, the uncertainty measure exhibits oscillatory dynamics in time, while in the {{spontaneously}} broken-symmetry phase and at exceptional points it evolves towards the case of a minimum-uncertainty steady state. We further showed that, in the {{spontaneously}} broken-symmetry regime, the eigenstates of the Hamiltonian are no longer orthogonal with respect to the metric-induced inner product, in agreement with recent experimental observations (see Eq.~(\ref{eq:over})).

Finally, by comparing our metric-based formulation with a Lindblad master-equation approach, we demonstrated that both descriptions yield the same steady-state behavior. This agreement provides an independent validation of our construction and clarifies the connection between non-Hermitian effective dynamics and open-system descriptions.

Our results highlight the necessity of incorporating appropriate metric operators —particularly beyond the exact symmetry phase— to extract reliable dynamical and statistical information from non-Hermitian Hamiltonians.
The framework developed here will be useful for 
a proper contextualization within the broader field of uncertainty relations and nonclassicality. 
Work is currently in progress to extend our analysis to the study of the Cramér–Rao bound in non-Hermitian systems \cite{cramerrao}, as well as to the investigation of stronger uncertainty relations \cite{maccone}.

\begin{acknowledgments}

The authors have been partially supported by projects PIP 0135 and PIP 0457 from the National Research Council CONICET (Argentina), and grants 11/X959 and 11/X982 from the National University of La Plata UNLP (Argentina). 
M.P., R.R. and M.R. are researchers of CONICET; Y.A. has a fellowship from CONICET.

\end{acknowledgments}

\MR{
\appendix
\section{ Results for $N$-particle systems.}\label{app}

In what follows, we present the results obtained for systems with $N$-levels. 

The spectrum of the Hamiltonian \ref{hamieq1} can be computed exactly. For $N$ odd, $N=2 n+1$ ($n=0,1,2,...$), the eigenvalues can be written as 
\begin{eqnarray}
\{\rho \pm (2 n +1) \lambda,\rho \pm (2 n -1) \lambda, ...,\rho \pm  \lambda\},\nonumber \\
\end{eqnarray}
and for $N$ even, $N=2 n$ ($n=0,1,2,...$), the eigenvalues can be exppressed as

\begin{eqnarray}
\{\rho \pm (2 n )  \lambda,\rho \pm (2 n -2) s  \lambda,...,\rho \pm 2  \lambda, 0\},\nonumber \\
\end{eqnarray}
with $\rho=r \cos(\theta)$, $\lambda= s \sqrt{1-\eta^2} $ and  $\eta= (r/s) \sin(\theta)$.

The EPs correspond to $\eta=\pm 1$ as for the two-level system.

We shall take as an initial state the Coherent Spin State (CSS) given by

\begin{eqnarray}
| z \rangle =\frac{1}{(1+|z|^2)^{\frac N 2}}~\sum_{k=0}^{N}~z^k \left( 
\begin{array}{c}
N\\
k
\end{array}
\right)^{\frac 12 } |k \rangle,
\end{eqnarray}
being $z= \tan(\theta_0/2) ~ {\rm e}^{-{\bf i} \phi_0}$, ${\bf n}(\theta_0,\phi_0) . {\bf S}= S$, with $S=N/2$.
The basis kets are given by $|k\rangle ={\cal N}_k~ S_+^k |S ~-S\rangle$, being  ${\cal N}_k$ the corresponding normalization constant.

The operators $\{S_x,S_y,S_z \}$ are the components of the collective spin of a system with total spin $S=N/2$, ${\bf S}={\mathbf {\sigma}}/2$.
We shall analyze uncertainty relations in the plane perpendicular to the mean value of spin at each instant of time. We indicate with $z'$ the direction of the mean value of the spin at each instant of time. In terms of the angles 
$(\theta_S, \phi_S)$, the direction of the component $S_{z'}$ is written as ${\bf n}_{z'}=(\sin(\theta_S) \cos(\phi_S),~\sin(\theta_S) \sin(\phi_S),~\cos(\theta_S))$. The perpendicular plane to $\langle {\bf S} \rangle$  is given by   $(x',~y')$. The direction for the component of spin with minimum uncertainty is $x'$. The squeezing parameter relates the fluctuation of the spin in a given direction to its mean value 

\begin{equation}
\zeta^2_{k'}=\frac {\Delta^2 S_{k'}} {\frac 12 |\langle S_{z'} \rangle |},~~~k'=x',y'.
\end{equation}
In terms of $\zeta^2_{k'}$, the uncertainty difference for the components of spin in the $x'$ and $y'$ direction is given by

\begin{equation}
UR= \left(\zeta^2_{x'} \zeta^2_{y'}-1 \right) \frac 14 |\langle S_{z'} \rangle |^2.
\end{equation}

In Figures \ref{figap1}, we plot the results obtained for the squeezing parameter and the orientation angles of the mean value of the spin as a function of the time, for a systems of $N=30$. Panels (a), (c), and (e) depict the squeezing parameter, $\zeta^2_{k'}$, while panels (b), (d), and (f) show the behavior of the orientation angles of the mean spin, $(\theta_S,\phi_S)$. Panels (a) and (b) correspond to $\eta=4$, panels (c) and (d) to $\eta=1/2$, and panels (e) and (f) to $\eta=1$. We have fixed $s=1/\sqrt{N}$ and $\theta=\pi/2$. The initial state is a CSS with $(\theta_0,\phi_0)=(\pi/4,0)$. With orange curves we plot $\zeta^2_{y'}$ and $\theta_S$, while with blue ones we depict $\zeta^2_{x'}$ and $\phi_S$. For values of $\eta>1$, the system is in the spontaneous broken phase. As shown in Panels (a) and (b) of both Figures, the systems reach and stationary state. For values of $\eta<1$, the system is in the unbroken phase. As can be observed in Panels (c) and (d) of both Figures, the squeezing parameter and the angles of the direction of mean value of spin show an oscillatory behaviour. The EP correspond to $\eta=\pm 1$. Panels (d) and (f) shows the behaviour of the observables at $\eta=1$. At this point the system reaches the minimum value for the uncertainties in both directions, $x'$ and $y'$. 

Figures \ref{figap2} show contour plots for the squeezing parameter as a function of the orientation of the initial CSS, $(\theta_0,\phi_0)$, for the system of $N=30$. The interaction parameters are set to $\eta=4$, $s=1/\sqrt{N}$, and $\theta=\pi/2$. Panels (a) and (c) display the behavior of $\zeta^2_{x'}$, while panels (b) and (d) show the behavior of $\zeta^2_{y'}$. Panels (a) and (b) correspond to $t=0$, whereas panels (c) and (d) correspond to $t=6$, after the system reaches its steady state. In Panel (a), red curves correspond to values of $\zeta^2_{x'}=1$, yellow curves correspond to values of $\zeta^2_{x'}<1$. As discussed before and shown in panels (c) and (d), the system evolves into a steady state with $\zeta^2_{x'} \approx 0.79$ and $\zeta^2_{y'} \approx 1.26$.

\begin{figure}[h!]
     \centering
     \includegraphics[width=1.0\linewidth]{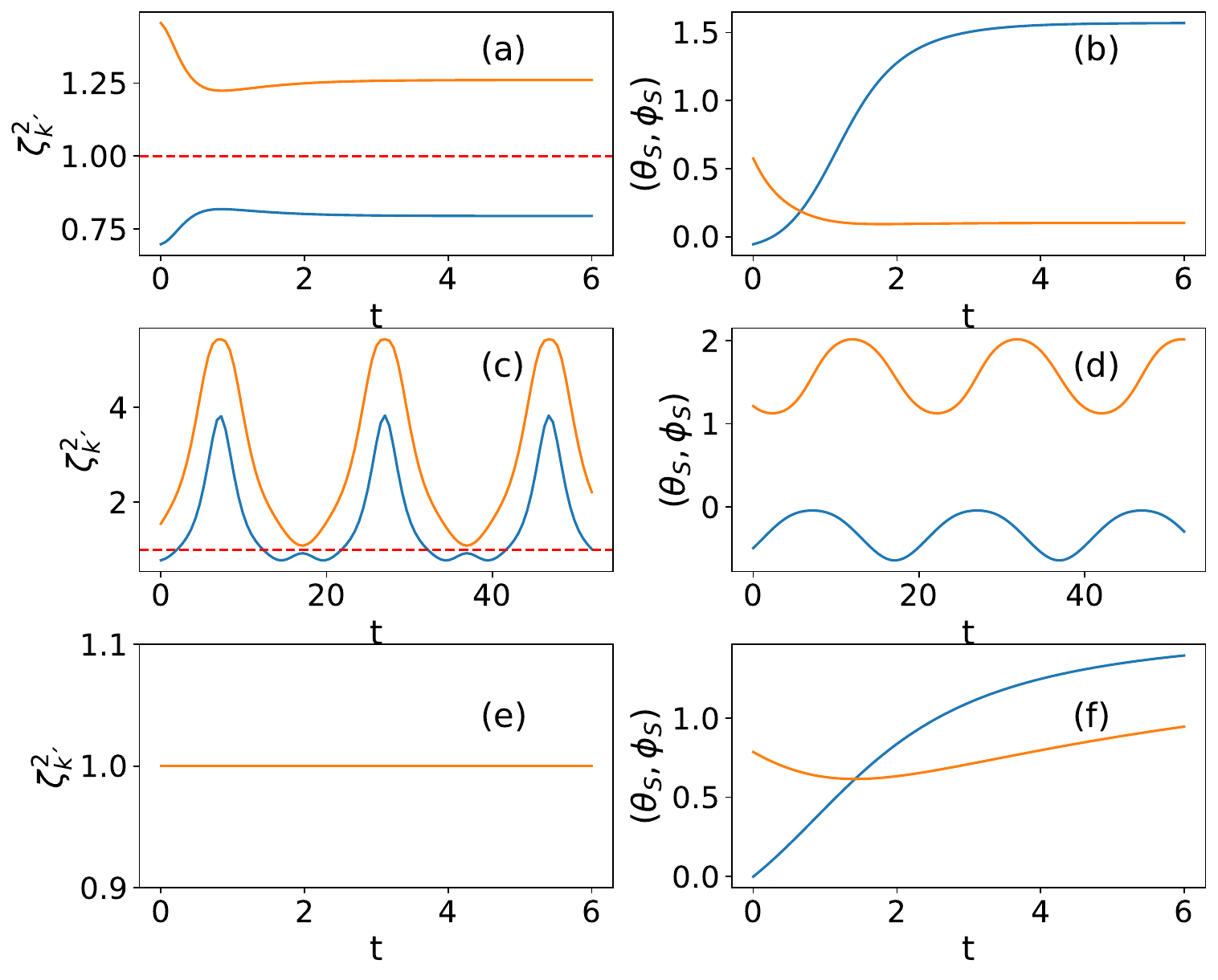}
     \caption{\MR{Panels (a), (c), and (e) depict the squeezing parameters $\zeta^2_{x'}$ (blue lines) and $\zeta^2_{y'}$ (orange lines), while panels (b), (d), and (f) show the behavior of the orientation angles of the mean spin, $\theta_S$ (in orange) and $\phi_S$ (in blue), as a function of time. The system consists of $N=30$ spins. Panels (a)-(b) correspond to $\eta=4$, panels (c)-(d) to $\eta=1/2$, and panels (e)-(f) to $\eta=1$. The Hamiltonian parameters are chosen as $s=1/\sqrt{N}$ and $\theta=\pi/2$. The initial state is a CSS with $(\theta_0,\phi_0)=(\pi/4,0)$.}}
     \label{figap1}
\end{figure}

\begin{figure*}[!htb]
     \centering
     \includegraphics[width=1.0\linewidth]{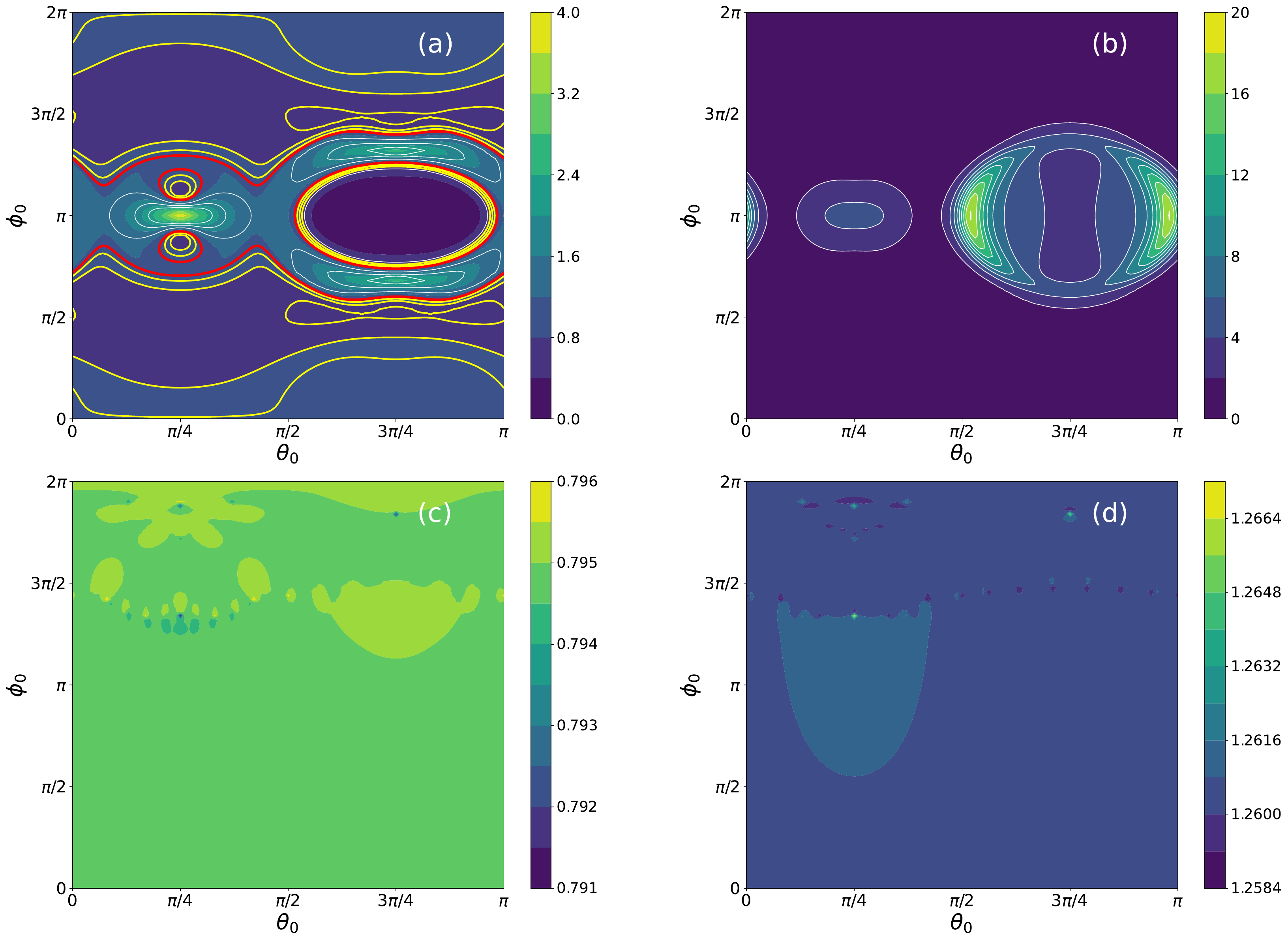}
     \caption{\MR{Contour plots for the squeezing parameter as a function of the orientation of the initial CSS, $(\theta_0,\phi_0)$. We consider a system of $N=30$ spins. The interaction parameters are set to $\eta=4$, $s=1/\sqrt{N}$, and $\theta=\pi/2$. Panels (a) and (c) display the behavior of $\zeta^2_{x'}$, while panels (b) and (d) show the behavior of $\zeta^2_{y'}$. Upper panels (a)-(b) correspond to $t=0$, whereas lower panels (c)-(d) correspond to $t=6$, after the system reaches its steady state. In panel (a), red curves correspond to values of $\zeta^2_{x'}=1$, yellow curves correspond to values of $\zeta^2_{x'}<1$.}}
     \label{figap2}
\end{figure*}

Finally, though there exists a preference direction given by the direction of the mean value of the spin at each instant of time, the generalized Heisenberg-Robertson uncertainty relation should remain valid when taking the components of ${\bf S}$ with respect to the original axes, ${\bf S}=(S_x, S_y, S_z)$. 

Figures \ref{figap3} displays results for the UR of Eq. (\ref{eq:ModRobertson}), written in terms of ${\sigma_x,\sigma_y,\sigma_z}$, as a function of time, for a systems with $N=30$. The Hamiltonian parameters have been fixed to values of $\eta=~4,~0.5,~1$, for insets (a), (b) and (c), respectively. The Hamiltonian parameters were fixed to $\theta=\pi/2$ and $s=1/\sqrt{N}$. We have considered an initial state with $\theta_0=\pi/4$ and $\phi_0=0$. As expected, the behavior of UR is oscillatory in the unbroken symmetry phase, it tends to a stationary value in the spontaneously broken symmetry phase and for the EP.

\begin{figure}[!h]
     \centering
     \includegraphics[width=1.0\linewidth]{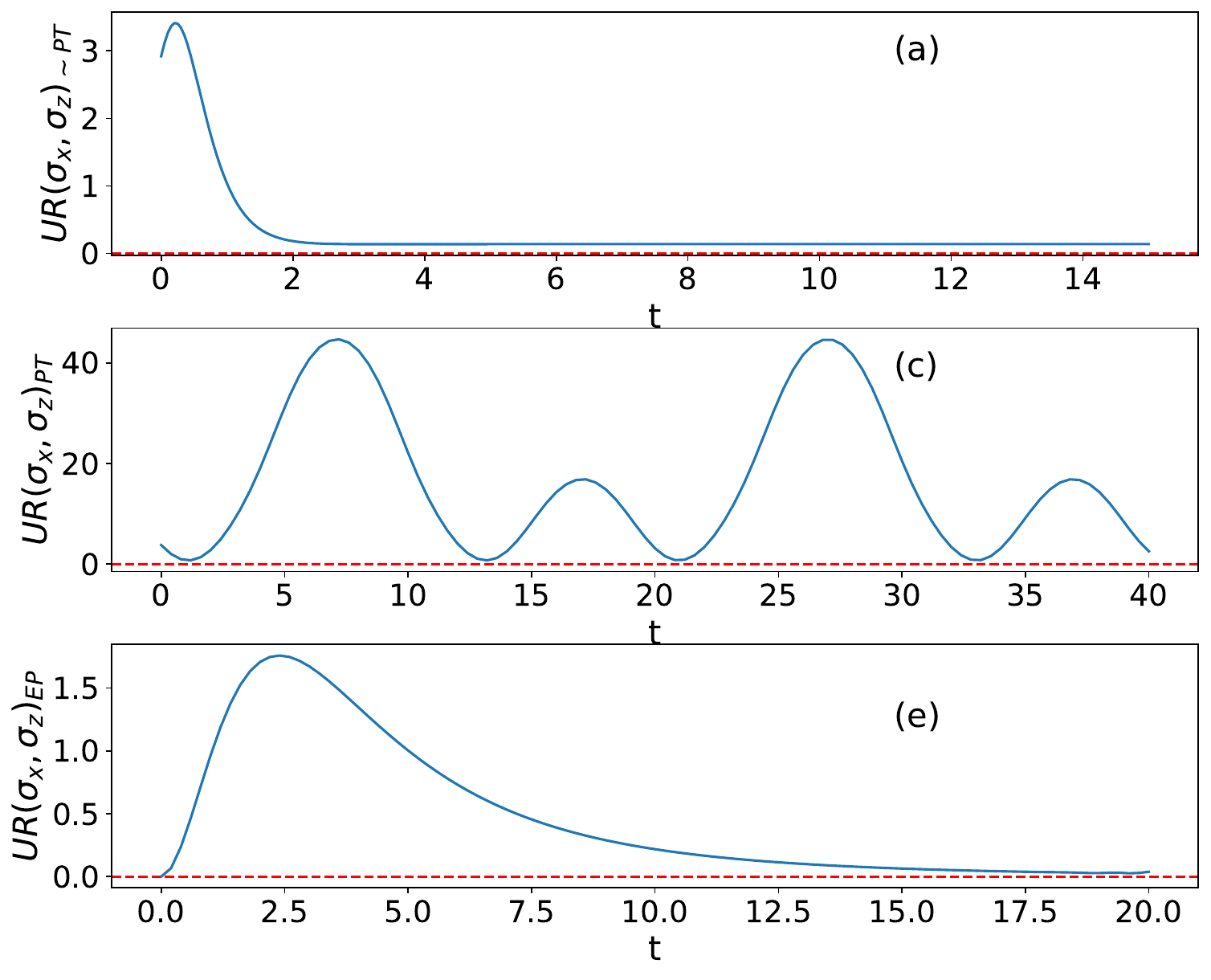}
     \caption{\MR{UR of Eq.(\ref{eq:ModRobertson}), written in terms of ${\sigma_x,\sigma_y,\sigma_z}$, as a function of 		time, for a systems with $N=30$. The Hamiltonian parameters have been fixed to values of $\eta=~0.5,~2,~1$, for insets (a), (b) and (c), respectively, $\theta=\pi/2$ and $s=1/\sqrt{N}$. We have considered an initial state with $\theta_0=\pi/4$ and $\phi_0=0$. 
}}
\label{figap3}
\end{figure}

Figure \ref{figap4} shows contour plots for the UR relation as a function of the orientation of the initial CSS, 
$(\theta_0,\phi_0)$. We consider a system of $N=30$. The interaction parameters are set $s=1/\sqrt{N}$, and $\theta=\pi/2$, and $\eta= 4$ for panels (a), (c) and (d), while $\eta=1$ for panels (b), (d) and (f). Panels (a)-(b) display results for $t=0$, panels (c)-(d) for t=3, and panels (e)-(f) for $t=6$. It is observed that as $t$ is incresed the UR tends to a stanionary value.

 \begin{figure*}[!htb]
     \centering
     \includegraphics[width=1.0\linewidth]{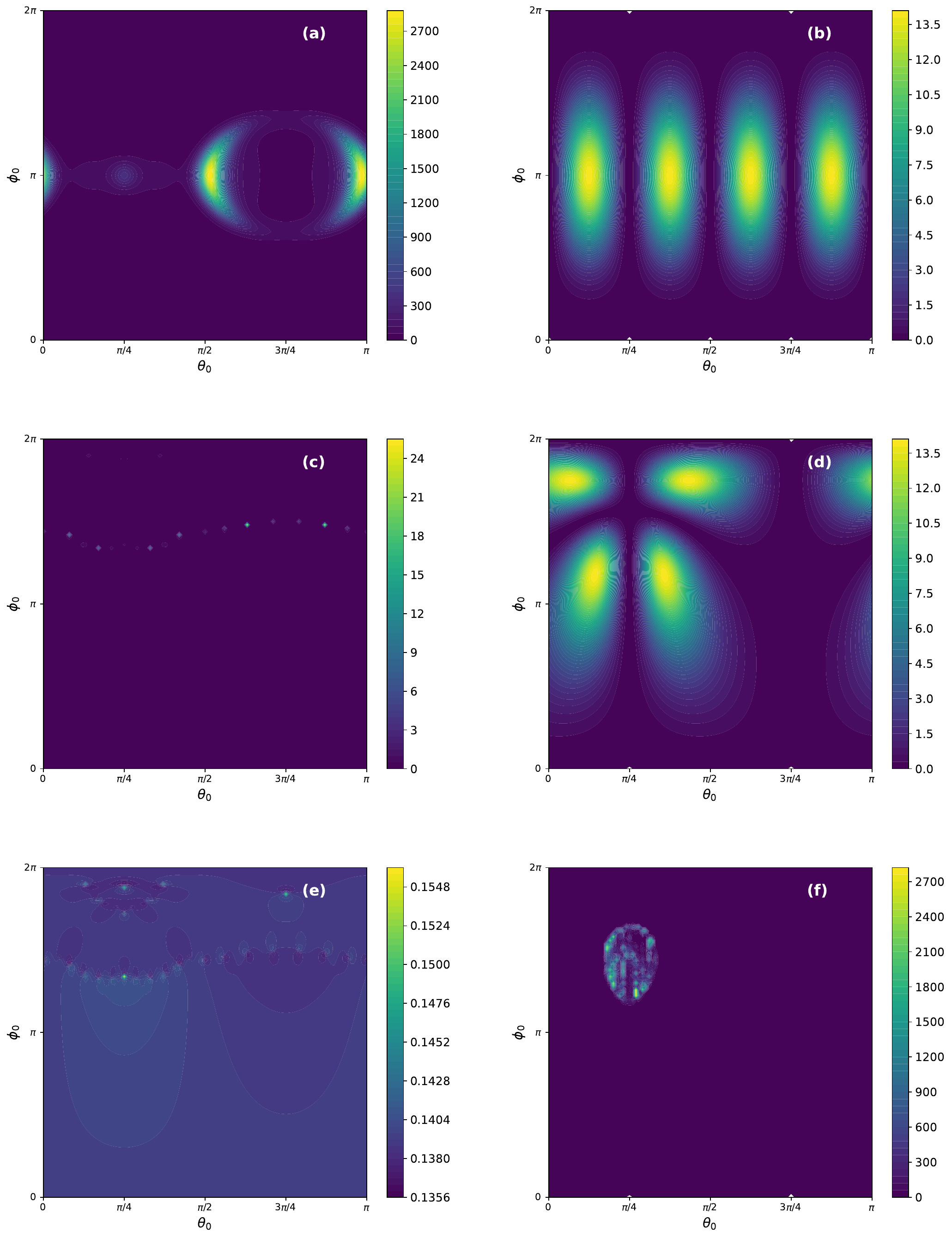}
     \caption{\MR{Contour plots for the URThe interaction parameters are set $s=1/\sqrt{N}$, and $\theta=\pi/2$, and $\eta= 4$ for panels (a), (c) and (d), while $\eta=1$ for panels (b), (d) and (f). Panels (a) and (b) display results for $t=0$, panels (c) and (d) for t=3, and panels (e) and (f) for $t=6$.  Red lines denotes the curve with $UR=0$.
}}
     \label{figap4}
\end{figure*}

}


%

\end{document}